%% file: BPH-21-002_temp.tex
\pdfoutput=1
\documentclass[11pt,twoside,a4paper,cmspaper,final,collab]{cms-tdr}
\def\svnVersion{693db24}\def\svnDate{2025/03/28}\def\cmsCernNoTag{CERN-EP-2024-268}\def\cmsCernDate{\today}\def\cmsMessage{Published in Physics Letters B as \href{http://dx.doi.org/10.1016/j.physletb.2025.139406}{\doi{10.1016/j.physletb.2025.139406}.}}
\begin{document}\cmsNoteHeader{BPH-21-002}

\newcommand{\qsq}{\ensuremath{q^{2}}\xspace}
\newcommand{\PHI}{\ensuremath{\phi}\xspace}
\newcommand{\thetak}{\ensuremath{\theta_{\PK}}\xspace}
\newcommand{\costl}{\ensuremath{\cos\theta_l}\xspace}
\newcommand{\costk}{\ensuremath{\cos\thetak}\xspace}
\newcommand{\costksq}{\ensuremath{\cos^2\thetak}\xspace}
\newcommand{\sintl}{\ensuremath{\sin\theta_l}\xspace}
\newcommand{\sintk}{\ensuremath{\sin\thetak}\xspace}
\newcommand{\sintlsq}{\ensuremath{\sin^2\theta_l}\xspace}
\newcommand{\sintksq}{\ensuremath{\sin^2\thetak}\xspace}
\newcommand{\CP}{\textit{CP}\xspace}

\DeclareRobustCommand{\kst}{{\HepParticle{\PK}{}{\ast0}}\Xspace}
\DeclareRobustCommand{\PKsti}{{\HepParticle{\PKst(892)}{}{0}}\Xspace}
\DeclareRobustCommand{\PBzalt}{{\HepParticle{\PB}{alt}{0}}\Xspace}
\DeclareRobustCommand{\PKstalt}{{\HepParticle{\PK}{alt}{\ast0}}\Xspace}

\newcommand{\bsll}{\ensuremath{\PQb\to\PQs\Pellp\Pellm}\xspace}

\newcommand{\psit}{\ensuremath{\PGyP{2S}}\xspace}
\newcommand{\bzKstarmassmm}{\ensuremath{\PBz\to\PKsti\Pgmp\Pgmm}\xspace}
\newcommand{\bzKstarmm}{\ensuremath{\PBz\to \kst\Pgmp\Pgmm}\xspace}
\newcommand{\bzKstarjpsi}{\ensuremath{\PBz\to \PJGy \kst}\xspace}
\newcommand{\bzKstarpsi}{\ensuremath{\PBz\to \psit \kst}\xspace}
\newcommand{\bzKstarmmext}{\ensuremath{\PBz\to \kst(\PKp\PGpm)\Pgmp\Pgmm}\xspace}
\newcommand{\kstmm}{\ensuremath{\PKp\PGpm\Pgmp\Pgmm}\xspace}

\newcommand{\fl}{\ensuremath{F_\mathrm{L}}\xspace}
\newcommand{\fs}{\ensuremath{F_S}\xspace}

\newcommand{\pone}{\ensuremath{P_1}\xspace}
\newcommand{\ptwo}{\ensuremath{P_2}\xspace}
\newcommand{\pthree}{\ensuremath{P_3}\xspace}
\newcommand{\pfourp}{\ensuremath{P^{\prime}_4}\xspace}
\newcommand{\pfivep}{\ensuremath{P^{\prime}_5}\xspace}
\newcommand{\psixp}{\ensuremath{P^{\prime}_6}\xspace}
\newcommand{\peightp}{\ensuremath{P^{\prime}_8}\xspace}

\newcommand{\scpdf}{\ensuremath{S^\mathrm{C}}\xspace}
\newcommand{\smpdf}{\ensuremath{S^\mathrm{M}}\xspace}
\newcommand{\epsc}{\ensuremath{\epsilon^{\mathrm{C}}}\xspace}
\newcommand{\epsm}{\ensuremath{\epsilon^{\mathrm{M}}}\xspace}
\newcommand{\Yb}{\ensuremath{Y_\mathrm{B}}\xspace}
\newcommand{\Ys}{\ensuremath{Y_\mathrm{S}}\xspace}
\newcommand{\Bang}{\ensuremath{B^\mathrm{a}}\xspace}
\newcommand{\Sang}{\ensuremath{S^\mathrm{a}}\xspace}
\newcommand{\Bmass}{\ensuremath{B^{m}}\xspace}

\newcommand{\pdf}{PDF\xspace}
\newcommand{\splot}{\ensuremath{{}_\mathrm{{s}}\mathcal{P}\mathrm{lot}}\xspace}

\newcommand\Tstrut{\rule{0pt}{2.6ex}} 
\newcommand{\ph}{\phantom{-}}
\newcommand{\phn}{\phantom{0}}
\ifthenelse{\boolean{cms@external}}{\providecommand{\figtwowidth}{0.32\textwidth}}{\providecommand{\figtwowidth}{0.35\textwidth}} 
\ifthenelse{\boolean{cms@external}}{\providecommand{\figfourwidth}{0.22\textwidth}}{\providecommand{\figfourwidth}{0.24\textwidth}} 

\cmsNoteHeader{BPH-21-002}
\title{Angular analysis of the \texorpdfstring{\bzKstarmassmm}{B0 to K*(892)0 mu+ mu-} decay in proton-proton collisions at \texorpdfstring{$\sqrt{s}= 13\TeV$}{sqrt(s) = 13 TeV}}

\author*[cern]{The CMS Collaboration}

\date{\today}

\abstract{
A full set of optimized observables is measured in an angular analysis of the decay \bzKstarmassmm using a sample of proton-proton collisions at $\sqrt{s}= 13\TeV$, collected with the CMS detector at the LHC, corresponding to an integrated luminosity of 140\fbinv. The analysis is performed in six bins of the squared invariant mass of the dimuon system, \qsq, over the range $1.1 < \qsq < 16\GeV^2$. The results are among the most precise experimental measurements of the angular observables for this decay and are compared to a variety of predictions based on the standard model. Some of these predictions exhibit tension with the measurements.
}

\hypersetup{
pdfauthor={CMS Collaboration},
pdftitle={Angular analysis of the B0 to K*(892)0 mu+ mu- decay in proton-proton collisions at sqrt(s) = 13 TeV},
pdfsubject={CMS},
pdfkeywords={CMS, B-physics, heavy flavor, rare decays, flavor anomalies}} 

\maketitle

\section{Introduction}
The \bsll flavor-changing neutral-current processes are strongly suppressed according to the standard model (SM) of particle physics, since they only proceed through electroweak loop-level diagrams. Since virtual contributions of physics beyond the SM could occur with comparable amplitudes and alter the decay properties, these transitions are an excellent environment for indirect new physics searches. Among these processes, the \bzKstarmmext decay, where \kst indicates the \PKsti meson, has long been considered a promising channel, thanks to its rich kinematical structure defined by the angular momenta of the final state particles. Compared to other well known decay modes, as for example $\PBzs\to\Pgmp\Pgmm$, where only the branching ratio and mean life are measured~\cite{ATLAS:2018cur,LHCb:2021awg,LHCb:2021vsc,CMS:2022mgd}, it provides access to a larger set of observables with high sensitivity to deviations from SM expectations.
An extensive program of measuring angular observables from this decay has been carried out by the BaBar~\cite{BaBar:2008fao}, Belle~\cite{Belle:2009zue,Wehle:2016yoi}, and CDF~\cite{CDF:2011tds} Collaborations.
The data collected during Run 1 of the CERN LHC (2011--2012) were used by the LHCb~\cite{LHCb:2013zuf,LHCb:2013ghj,Aaij:2015oid}, CMS~\cite{CMS:2013mkz,CMS:2015bcy,Sirunyan:2017dhj}, and ATLAS~\cite{ATLAS:2018gqc} Collaborations to investigate this decay mode. 
Since the outcomes of those measurements were statistically limited, the additional data collected during Run 2 (2015--2018) can be used to significantly reduce the experimental uncertainties. 
The most recent model-independent result from the LHCb Collaboration~\cite{Aaij:2020nrf}, as well as two alternative analyses~\cite{LHCb:2023gpo,LHCb:2024onj}, confirm their previous measurement and appear to exhibit tension with some of the predictions based on the SM.
These predictions, while all based on the SM, yield different values with varying uncertainties. 
The largest differences in the calculations arise from the estimation of the non-local hadronic matrix elements and the evaluation of the corresponding uncertainty~\cite{Ciuchini:2021smi,Alguero:2023jeh,Bordone:2024hui}. 

This Letter contains the first CMS measurement of a complete set of optimized observables averaged over charge-conjugation and parity (\CP) of the \bzKstarmm decay. The results are obtained from the Run 2 data set collected in 2016--2018 at a center-of-mass energy of 13\TeV, corresponding to an integrated luminosity of 140\fbinv.
The analysis is performed in bins of the invariant mass squared of the dimuon system, \qsq, ranging from 1.1 to $16\GeV^2$. 
The previous CMS analyses of this decay mode utilized smaller data sets from $\sqrt{s} = 7--8$\TeV collisions and measured only a partial set of the angular observables.
Tabulated results are provided in the HEPData record for this analysis~\cite{hepdata}.
Throughout this Letter, charge-conjugate states are implied unless explicitly stated otherwise. 

\section{Decay rate and angular analysis observables}
\label{sec:formula}

\begin{figure*}[hbtp]
  \centering
  \includegraphics[width=0.8\textwidth]{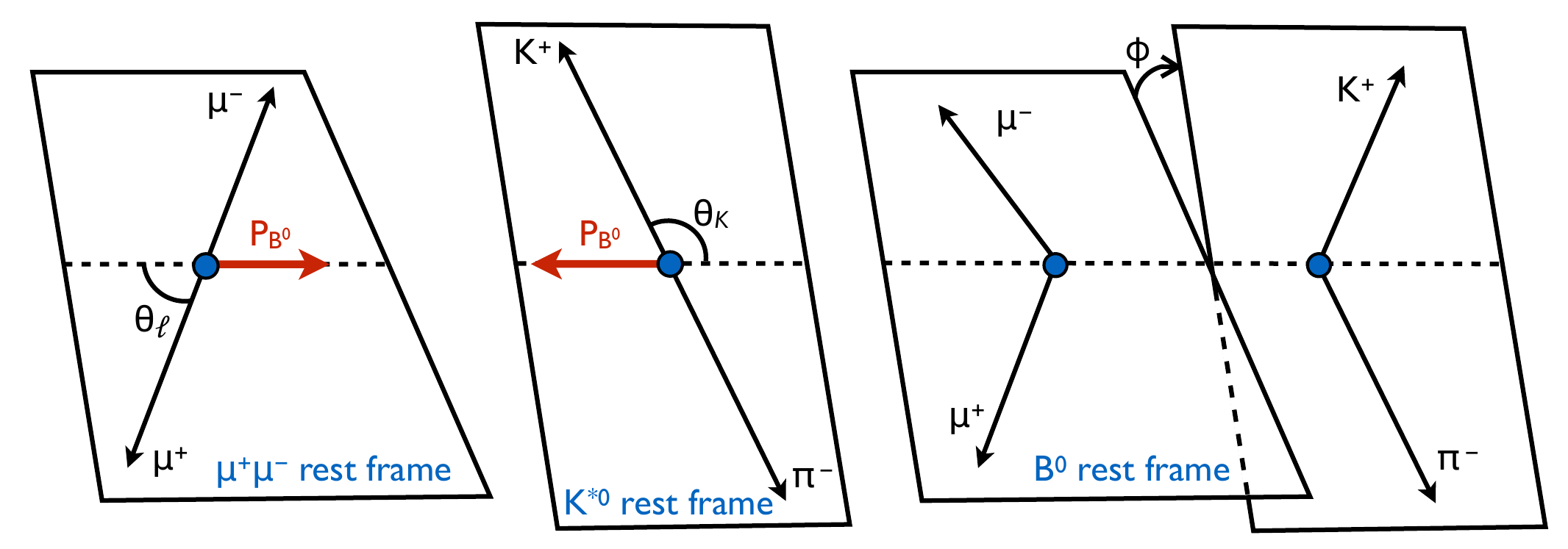}
  \caption{Sketch representing the definition of the angles $\theta_l$ (left), \thetak (center), and \PHI (right).}
  \label{fig:anglesDef}
\end{figure*}

The decay \bzKstarmm, where the \kst meson decays to $\PKp\PGpm$, can be described by \qsq and three decay angles: $\theta_l$, \thetak, and \PHI.
The definition of the angles is consistent with the one used in experimental publications and detailed in Ref.~\cite{LHCb:2013zuf}. 
A sketch representing the definition of the angles is shown in Fig.~\ref{fig:anglesDef}.
The angle $\theta_l$ is defined as the angle between the positive (negative) muon momentum and the direction opposite to the \PBz\ $\big(\PABz\big)$ momentum in the dimuon rest frame;
the angle \thetak is defined as the angle between the kaon momentum and the direction opposite to the \PBz\ $\big(\PABz\big)$ momentum in the \kst $\big(\PAKst{}^{0}\big)$ rest frame;
the angle \PHI is the angle between the plane containing the \Pgmp and \Pgmm and the plane containing the kaon and the pion from the \kst decay in the \PBz rest frame. 

Following the notation of Ref.~\cite{Altmannshofer:2008dz} and the approach further proposed in Refs.~\cite{DescotesGenon:2012zf,Descotes-Genon:2013vna}, the \CP-averaged angular decay rate of the \bzKstarmm decay can be written as:
\begin{equation} \label{eq:decayPi}
\ifthenelse{\boolean{cms@external}}
{ 
    \begin{split}
  \Gamma_P\equiv\frac{1}{\rd\Gamma/\rd\qsq} & \frac{\rd^4\Gamma}{\rd\qsq \rd\mathrm{cos}\theta_l \rd\mathrm{cos}\thetak \rd\phi} = \\ 
         & \frac{9}{32\pi}  \Bigg[ \frac{3}{4}  (1-\fl)\sintksq + \fl \costksq \\
		 & + \Bigg( \frac{1}{4} (1-\fl)\sintksq - \fl \costksq \Bigg) \cos{2\theta_l}\\
         & + \frac{1}{2} \pone (1-\fl)\sintksq \sintlsq \cos{2\phi} \\
         & + \sqrt{(1-\fl) \fl} \, \Bigg(\frac{1}{2} \pfourp \sin{2\thetak}\sin{2\theta_l}\cos{\phi} \\
         & + \pfivep \sin{2\thetak} \sintl\cos{\phi}\Bigg) \\
         & - \sqrt{(1-\fl) \fl} \, \Bigg(\psixp \sin{2\thetak}\sintl\sin{\phi} \\
         & + \frac{1}{2} \peightp \sin{2\thetak} \sin{2\theta_l}\sin{\phi}\Bigg) \\
		 & + 2 \ptwo (1-\fl) \sintksq \costl \\
		 & - \pthree (1-\fl) \sintksq \sintlsq \sin{2\phi} \Bigg],
    \end{split}
}
{ 
    \begin{split}
  \Gamma_P\equiv\frac{1}{\rd\Gamma/\rd\qsq} & \frac{\rd^4\Gamma}{\rd\qsq \rd\mathrm{cos}\theta_l \rd\mathrm{cos}\thetak \rd\phi} = \frac{9}{32\pi} \left[ \frac{3}{4} (1-\fl)\sintksq + \fl \costksq \right. \\
		 & + \left(\frac{1}{4} (1-\fl)\sintksq - \fl \costksq \right) \cos{2\theta_l}\\
         & +  \frac{1}{2} \pone (1-\fl)\sintksq \sintlsq \cos{2\phi} \\
         & + \sqrt{(1-\fl) \fl}\left(\frac{1}{2} \pfourp \sin{2\thetak}\sin{2\theta_l}\cos{\phi} +  \pfivep \sin{2\thetak} \sintl\cos{\phi}\right) \\
         & - \sqrt{(1-\fl) \fl}\left(\psixp \sin{2\thetak}\sintl\sin{\phi}  + \frac{1}{2} \peightp \sin{2\thetak} \sin{2\theta_l}\sin{\phi}\right) \\
		 & \left. + 2 \ptwo (1-\fl) \sintksq \costl - \pthree (1-\fl) \sintksq \sintlsq \sin{2\phi} \right],
    \end{split}
}         
\end{equation}
where \fl is the fraction of longitudinal polarization of the \kst meson, and $P_{1,2,3}$ and $P'_{4,5,6,8}$ are observables that have been optimized to reduce their dependence on the values and uncertainties of the form factors~\cite{DescotesGenon:2012zf}.
These observables are functions of the decay amplitudes, which themselves depend on the Wilson coefficients contained in the relevant effective Hamiltonian~\cite{Altmannshofer:2008dz}. 
In the publications of the LHCb Collaboration~\cite{Aaij:2015oid,Aaij:2020nrf} the definition of some of the $P_i^{(\prime)}$ observables is different, namely $\pfourp{}^{\mathrm{CMS}} = 2 \pfourp{}^{\mathrm{LHCb}}$, $\psixp{}^{\mathrm{CMS}} = -\psixp{}^{\mathrm{LHCb}}$, and $\peightp{}^{\mathrm{CMS}} = -2 \peightp{}^{\mathrm{LHCb}}$.

In order to also describe $\PBz\to \PKp\PGpm\Pgmp\Pgmm$ events in which the $\PKp\PGpm$ system can be in an $S$-wave instead of $P$-wave configuration, the following extension of the angular decay rate (as in Eq.~(7.8) of Ref.~\cite{Descotes-Genon:2013vna}) is used:
\begin{equation} \label{eq:decaySw}
\ifthenelse{\boolean{cms@external}}
{ 
    \begin{split}
    \frac{1}{\rd\Gamma/\rd\qsq} & \frac{\rd^4\Gamma}{\rd\qsq \rd\mathrm{cos}\theta_l \rd\mathrm{cos}\thetak \rd\phi} = (1-\fs) \Gamma_P \\
    & + \frac{3}{16\pi}  ( \fs \sintlsq + A_S  \sintlsq \costk \\ 
    & + A^4_S \sintk \sin{2\theta_l} \cos{\phi} + A^5_S \sintk \sintl \cos{\phi} \\ 
    & + A^7_S \sintk \sintl \sin{\phi} + A^8_S \sintk \sin{2\theta_l} \sin{\phi} ),
  \end{split}
}
{ 
  \begin{split}
    \frac{1}{\rd\Gamma/\rd\qsq} & \frac{\rd^4\Gamma}{\rd\qsq \rd\mathrm{cos}\theta_l \rd\mathrm{cos}\thetak \rd\phi} = (1-\fs) \Gamma_P + \frac{3}{16\pi}  ( \fs \sintlsq +
    A_S  \sintlsq \costk \\ & +
    A^4_S \sintk \sin{2\theta_l} \cos{\phi} +
    A^5_S \sintk \sintl \cos{\phi} \\ & +
    A^7_S \sintk \sintl \sin{\phi} +
    A^8_S \sintk \sin{2\theta_l} \sin{\phi} ),
  \end{split}
}
\end{equation}
where the term proportional to \fs is the $S$-wave component, and the terms proportional to the $A_S$ and $A^{4,5,7,8}_S$ coefficients describe the interference between the $S$- and $P$-wave terms.

\section{The CMS detector and event samples}
The central feature of the CMS apparatus is a superconducting solenoid of 6\unit{m} internal diameter, providing a magnetic field of 3.8\unit{T}. Within the solenoid volume are a silicon pixel and strip tracker, a lead tungstate crystal electromagnetic calorimeter, and a brass and scintillator hadron calorimeter, each composed of a barrel and two endcap sections. 
Forward calorimeters extend the pseudorapidity ($\eta$) coverage provided by the barrel and endcap detectors. 
Muons are measured in the range $\abs{\eta} < 2.4$, with detection planes made using three technologies: drift tubes, cathode strip chambers, and resistive-plate chambers. 
The single-muon trigger efficiency exceeds 90\% over the full $\eta$ range, and the efficiency to reconstruct and identify muons is greater than 96\%. 
Matching muons to tracks measured in the silicon tracker results in a relative transverse momentum (\pt) resolution, for muons with \pt up to 100\GeV, of 1\% in the barrel and 3\% in the endcaps~\cite{CMS:2018rym}. 
The silicon tracker used in 2016 measured charged particles within the range $\abs{\eta} < 2.5$. 
For nonisolated particles of $1 < \pt < 10\GeV$ and $\abs{\eta} < 1.4$, the track resolutions were typically 1.5\% in \pt and 25--90 (45--150)\mum in the transverse (longitudinal) impact parameter~\cite{CMS:2014pgm}. 
At the start of 2017, a new pixel detector was installed~\cite{Phase1Pixel}; the upgraded tracker measured particles up to $\abs{\eta} = 3.0$ with typical resolutions of 1.5\% in \pt and 20--75\mum in the transverse impact parameter~\cite{DP-2020-049} for nonisolated particles of $1 < \pt < 10\GeV$.
A more detailed description of the CMS detector, together with a definition of the coordinate system used and the relevant kinematic variables, can be found in Ref.~\cite{CMS:2008xjf}.

Events of interest are selected using a two-tiered trigger system. The first level, composed of custom hardware processors, uses information from the calorimeters and muon detectors to select events at a rate of around 100\unit{kHz} within a fixed latency of about 4\mus~\cite{CMS:2020cmk}. 
The second level, known as the high-level trigger, consists of a farm of processors running a version of the full event reconstruction software optimized for fast processing, and reduces the event rate to around 1\unit{kHz} before data storage~\cite{CMS:2016ngn}.
All events used in this analysis are selected by a set of triggers demanding two identified muons of opposite charges and an additional track to form a common vertex, displaced from the region of the proton-proton interactions (beamspot). The trigger required each muon to have $\pt>4\GeV$, $\abs{\eta} < 2.5$, and to pass within 2 cm of the beam axis. 
The dimuon system was required to have $\pt>6.9\GeV$, a vertex fit $\chi^2$ probability larger than 10\%, a separation of the vertex relative to the beamspot in the transverse plane larger than 3 times the uncertainty of the separation, and the cosine of the angle in the transverse plane between the dimuon momentum vector and the vector from the beamspot to the dimuon vertex greater than 0.9.
The additional track was required to have $\pt>0.8\,(1.2)\GeV$ and a minimum distance of closest approach from the beamspot greater than 0 (2) times the uncertainty for data collected in 2016 (2017--2018).
As the trigger configuration was restricted in the \qsq range, the analysis is limited to events with $\qsq<16\GeV^2$.

Monte Carlo (MC) simulations are used to determine the impact of the detector geometry, trigger, reconstruction, and selection on the angular distribution of the signal, and to validate the analysis procedure. 
Simulated events are generated with \PYTHIA 8.230~\cite{Sjostrand:2014zea}, decays of particles containing b or c quarks are simulated with \EVTGEN 1.6.0~\cite{Lange2001152}, and final-state radiation is included via \PHOTOS 3.61 ~\cite{BARBERIO1991115,BARBERIO1994291}.
Events are passed through the CMS detector simulation based on \GEANTfour~\cite{Agostinelli:2002hh} and simulated minimum bias events are added to mimic the number of multiple vertices per event as seen in the data.

\section{Event reconstruction, selection, and efficiency modeling}
The offline reconstruction requires two oppositely charged muons and two oppositely charged tracks.
Muons must satisfy general identification (soft muon) criteria~\cite{CMS:2018rym} and match the muons used in the trigger. 
The \pt and $\eta$ requirements applied by the trigger must be satisfied by the offline muons as well.
The two tracks are required to pass the standard high-quality~\cite{CMS:2014pgm} identification criteria, have $\abs{\eta}<2.4$ and $\pt > 0.8\GeV$ to ensure well-measured tracks and match the trigger requirements, and have a distance of closest approach to the beamspot greater than 0.8 times the uncertainty to preferentially select tracks originating from a displaced vertex.
One of the two tracks must match the track used in the trigger.
In the 2017 and 2018 data sets, the trigger-matched track is required to have $\pt > 1.2\GeV$ and the distance of the closest approach to the beamspot greater than twice the uncertainty.
A veto is applied to reject tracks that are also reconstructed and identified as muons.
The two tracks are fitted to a common vertex to form the \kst candidate.
The \kst meson invariant mass is computed for both the $\PKp\PGpm$ and $\PKm\PGpp$ mass hypotheses, and at least one of the two combinations is required to lie within 150\MeV of the nominal \kst meson mass~\cite{ParticleDataGroup:2024cfk}.
A \PGfP{1020} veto is applied, demanding that the invariant mass of the two-track system, when assigning the kaon mass to both, is larger than 1.035\GeV.

The \PBz candidates are formed by combining the two oppositely charged muons with a \kst candidate.
The four daughter tracks of the \PBz candidate must originate from a common vertex, with vertex fit $\chi^2$ probability greater than 1\%.
The momenta of the four tracks are then refitted using the common vertex as a constraint to improve the resolution of the track parameters.
The flavor of the \PBz and \kst candidates, and hence the mass assigned to the two hadronic tracks, is determined by the mass assignment that gives a \kst meson mass closest to its nominal value.
Based on simulation studies, this criterion results in incorrect flavor assignment for 12--14\% of the candidates, depending on the data-taking year and with an increasing trend as a function of \qsq.

A multivariate analysis is used to optimize the background rejection.
A boosted decision tree is trained on background data events from the mass sidebands and simulated signal events, separately for each year of data-taking.
The input variables of the discriminator are: the $\chi^2$ probability of the \PBz candidate decay vertex; the transverse distance between the \PBz candidate decay vertex and the beamspot divided by its
uncertainty; the cosine of the angle between the \PBz candidate momentum and the distance vector from the beamspot to the \PBz candidate decay vertex, in the transverse plane; the transverse \PBz candidate impact parameter 
with respect to the beamspot divided by its uncertainty; the transverse impact parameter of each hadronic track with respect to the beamspot divided by its uncertainty; the \kst candidate invariant mass; and the sum of the final-state particles relative isolation, defined as the sum of the \pt of all the high-quality~\cite{CMS:2014pgm} tracks having $\pt>0.8\GeV$ within a cone of 
$\Delta R = \sqrt{\smash[b]{(\Delta\eta)^2 + (\Delta\phi)^2}} < 0.4$ centered on the particle, divided by the \pt of the particle.
Candidates with a decision tree score above a threshold are retained for analysis, with the threshold set to optimize the expected signal significance.

To reduce the contamination from $\PBp\to\PKp\PGmp\PGmm$, including the dimuon decays of \PJGy and \psit mesons, a veto is imposed on the invariant mass of the two possible $\PKp\PGmp\PGmm$ triplets, where \PKp is, in turn, one of the two hadronic tracks from the \kstmm candidate, with the charged kaon mass assigned. 
The event is discarded if the mass of either $\PKp\PGmp\PGmm$ combination is within a region corresponding to three times the $\PKp\PGmp\PGmm$ mass resolution around the \PBp meson mass. 
This selection suppresses the contamination from these decays to a negligible level.

The \PJGy and \psit meson control samples are defined by the ranges of the dedicated \qsq bins, 8.68--$10.09\GeV^2$ and 12.86--$14.18\GeV^2$, and the requirement $\abs{q-m_{\PJGy}}<3\sigma_q$ and $\abs{q-m_{\psit}}<3\sigma_q$, respectively. Here $\sigma_q$ is the per-event uncertainty in the dimuon invariant mass $q$, and $m$ represents the nominal mass of the charmonium state.
Events satisfying these requirements are excluded from the signal \qsq bins.
However, a fraction of events from the resonant channels still appears in the adjoining \qsq bins, mainly due to an unreconstructed photon from the charmonium decay. 
This contamination is reduced by applying combined requirements on $q$ and the \PBz candidate invariant mass, which are tuned independently for the dimuon mass regions below the \PJGy meson mass, between the \PJGy and
the \psit meson masses, and above the \psit meson mass.
These requirements are set so that less than 10\% of the background events originate from the resonant channels, by using simulated events to measure the remaining resonant contamination.
For $q < m_{\PJGy}$ $(q > m_{\PJGy})$, candidates must satisfy the condition $\abs{(m - m_{\PBz}) - (q - m_{\PJGy})} > 190\,(90)\MeV$, where $m$ is the invariant mass of the \PBz candidate. 
For $q < m_{\psit}$ $(q > m_\psit)$, candidates must satisfy the condition $\abs{(m - m_{\PBz}) - (q - m_\psit)} > 70\,(80)\MeV$. 
These requirements are tightened to 200, 100, 80, and 110\MeV in the 2016 data sample, as a consequence of the different detector configuration, trigger conditions, and offline selection criteria.
In addition, \PBz candidates formed by the final-state particles from the \bzKstarjpsi decay, where the identification of a muon and a hadron track is inverted, 
are rejected with dedicated requirements obtained from simulation. Two alternative \PBz candidates, denoted as ``alt'', are formed by reversing the identification of the same-sign muon and hadron track pairs, and
the one giving a \kst mass closest to its nominal value is selected (\PBzalt). Requirements on $m(\PKstalt)$, $q_{\text{alt}}$, $m(\PBzalt)$, $\Delta R(\PGm,\text{track})$, and $\Delta \pt(\PGm,\text{track})$ 
are placed so that 90\% of the reversed candidates are rejected, with less than 1\% impact on the signal efficiency.
Peaking backgrounds from $\PBzs\to \PGmp\PGmm\PGfP{1020}$, $\PBzs\to \PGmp\PGmm \kst$ and $\PBzs\to \PGmp\PGmm \PKp\PKm$ decays are studied in simulation, using their equivalent decays with intermediate \PJGy and \psit resonances. 
After the full reconstruction and selection chain, their contribution is negligible compared to the \PJGy resonant channel decay.

The contamination from $\PGLzb\to \Pp\PKm\PGmp\PGmm$ decays is investigated directly with data by assigning the proton and kaon masses to the hadronic tracks of the \kstmm candidates, and searching for a peak at the nominal \PGLzb mass in the candidate invariant mass distribution.
No evidence of such contamination is found.

Finally, partially reconstructed decays $\PBp\to\PKp\psit(\to\PJGy\PGpp\PGpm)$ contribute to the low-mass sideband of the \PJGy control sample and present a narrow peak in the \costk distribution in the region $0.6<\costk<0.8$. 
To reject this background, a selection is applied on $m(\PGm\PGm\PK)$, $m(\PGm\PGm\PGp)$, $m(\PGm\PGm\PK_\PGp\PGp_\PK)$, and $m(\PK_\PGp\PGp_\PK)$ invariant masses in the events with $\pt(\PK)>\pt(\PGp)$, 
where $\PK_\PGp$ and $\PGp_\PK$ are given the kaon and pion flavor assignment, respectively, based on which combination is closest to the \kst mass, but are assigned the charged pion and kaon mass, respectively, in the invariant mass calculations.

The acceptance, as well as the efficiency to trigger, reconstruct, and select the signal and control channel events, are modeled with a function of the three angular variables, using simulated event samples, individually for each year of data-taking and \qsq bin.
These functions are obtained by applying the nonparametric kernel density estimator method~\cite{Cranmer:2000du,Scott2015-xp} to the unbinned distributions of simulated events before and after imposing the acceptance and selection criteria previously described, and computing their ratio.
Studies of this method using simulated events find that it reproduces important features of the efficiency and remains robust in regions of the phase space that are poorly populated by simulated events.
Efficiencies are determined for both correctly identified (the kaon and pion have the correct charge) and misidentified (the kaon and pion charges are reversed) candidates, with both contributing as signal in the fit described below.
Corrections derived from the data are applied to the simulated events to improve the modeling of the additional proton-proton interactions within the same or nearby bunch crossings and of those variables correlated with the decay angles. The latter are evaluated by training a classifier on the difference between simulation and data from the \bzKstarjpsi control sample.
The background-subtracted data distributions have been obtained via the \splot technique~\cite{Pivk:2004ty}, where the reconstructed \PBz candidate invariant mass is used as the discriminating variable. 
The inputs to the classifier are the transverse momenta of the final-state particles, the \PBz candidate $\eta$, and the topological variables that are correlated to the angular ones.
The output score of the classifier is used to reweight the simulation. The effectiveness of the method has been validated assessing the compatibility of the distributions of the input features in data and in the corrected simulation, on an independent sample of \bzKstarjpsi events, as well as on a sample of \bzKstarpsi events.

\section{Fit procedure}
The angular observables are measured in each \qsq bin by performing an unbinned maximum-likelihood fit to the four-dimensional distribution of the invariant mass of the \PBz candidate, $m$, and the three angular variables: \costk, \costl, and \PHI. The bin definition is the same as in Ref.~\cite{Sirunyan:2017dhj}, except that the first bin starts from $1.1\GeV^2$ rather than $1.0\GeV^2$ to exclude the contamination from the \PGfP{1020} resonance.
The exact bin definition is reported in Table~\ref{tab:results}.
All \PBz candidates with the invariant mass in the range $5 < m < 5.6\GeV$ are included and a simultaneous fit is performed to data sets collected in 2016, 2017, and 2018.

The probability density function (\pdf) used to fit each of these three data sets is:
\begin{equation} \label{eq:angALL}
\ifthenelse{\boolean{cms@external}}
{ 
  \begin{split}
     P(&m,\costk,\costl,\PHI) = \\
    & \Ys \biggl[ \scpdf(m) \, \Sang(\costk,\costl,\PHI) \, \epsc(\costk,\costl,\PHI) \biggr. \\
    & + R\smpdf(m) \\
    & \times\Sang(-\costk,-\costl,-\PHI)  \epsm(\costk,\costl,\PHI) \biggr] \\
    & + \Yb\Bmass(m) \Bang(\costk,\costl,\PHI),
  \end{split}
}
{
  \begin{split}
    P(m,\costk,\costl,\PHI) & = \Ys \biggl[ \scpdf(m)  \, \Sang(\costk,\costl,\PHI) \, \epsc(\costk,\costl,\PHI) \biggr. \\
      & \biggl. + \, R \, \smpdf(m) \, \Sang({-}\costk,{-}\costl,{-}\PHI) \, \epsm(\costk,\costl,\PHI) \biggr] \\
    & + \Yb\,\Bmass(m) \, \Bang(\costk,\costl,\PHI),
  \end{split}
}
\end{equation}
where the first and second terms represent the contributions of the signal events for which the flavor has been correctly and wrongly identified, respectively, while the third term describes the background contribution.
The \Ys and \Yb parameters represent the yield of signal and background events, respectively.
The functions \scpdf and \smpdf describe the mass distribution of correctly identified and misidentified signal candidates, respectively, and they are parametrized, depending on the \qsq bin, by a Crystal Ball
function~\cite{CrystalBallRef} plus a Gaussian or a double Crystal Ball function~\cite{CrystalBallDouble} (with a single or double Gaussian kernel and two power-law tails in opposite directions).
The function \Sang is the angular decay rate as written in Eq.~\eqref{eq:decaySw}.
The functions \epsc and \epsm represent the efficiency for correctly and misidentified events, respectively, as a function of the angular variables. 
The parameter $R$ represents the ratio of the relative amount of misidentified events in data and the same value as computed on MC samples; the information of the latter value is contained in the relative value of \epsm with respect to \epsc, while $R$ is introduced to account for differences between data and simulation.
The function \Bmass represents the mass distribution of the combinatorial background events and consists of an exponential function; in the ranges $6 < \qsq < 8.68\GeV^2$ and $10.09 < \qsq < 12.86\GeV^2$, the exponential is multiplied by a function describing the reduction of background events from the requirements used to reduce the \PJGy and \psit meson contamination.
The function \Bang represents the angular distribution of the background, which is modeled by a three-dimensional Bernstein polynomial determined from the data sidebands. 
The sidebands are regions of the \kstmm mass spectrum that do not contain the signal peak. Their range is defined to minimize the contamination from signal events, while providing a sufficiently large sample to ensure a good fit of the angular distribution. This approach results in signal contamination below 6\% for the events within the sideband regions. 
The Bernstein polynomial degree is chosen as the lowest value that produces a $p$-value larger than 5\%: the degree varies between 2 and 6 for \costk, between 1 and 5 for \costl, and between 2 and 5 for \PHI, depending on the data-taking year and \qsq bin.
The functions \epsc, \epsm, and \Bang are constrained to have the same value at $\PHI = \pm \pi$.

In the simultaneous fit, the parameters of the \pdf corresponding to the angular observables are shared among the three data sets, while other parameters are independent to account for the effects of different data-taking conditions on the efficiency and resolution. 
The coefficients of the $S$-wave and its interference with the $P$-wave are treated as nuisance parameters and are shared among the three data sets.
All parameters are free, except those of the background angular shape, which are fixed to the values obtained from the data sidebands as described above.
The parameters of the signal mass shapes are constrained to the results of the fit to the \kstmm mass distribution of simulated signal events through Gaussian likelihood factors.
An additional Gaussian constraint is applied to the parameter $R$, with a mean equal to one and a width corresponding to the expected statistical fluctuation of $R$ in the data, which varies from 5 to 20\% depending on the data-taking year and the \qsq bin.
The fit strategy comprises a procedure relying on the penalization of the likelihood function to ensure that the resulting angular observables lie in the physical region of the parameter space, \ie provide a non-negative decay rate.

The best fit value of the parameters and approximate 68\% \CL confidence intervals are extracted using scans of the profile likelihood ratio, following the procedure described in Section~3.2 of Ref.~\cite{Khachatryan:2014jba}.
The validity of this method and its robustness to the presence of nonphysical regions of the parameter space have been proven by an extensive study of its statistical coverage. 
The fit procedure and results are validated through fits to pseudo-experiments, MC simulation samples, and control samples.

\section{Systematic uncertainties}
Several sources of systematic uncertainties are evaluated to account for possible biases introduced by the analysis procedure, differences between the signal features in data and simulation, and any unaccounted minor background sources. 

The possible bias introduced by mismodeling the efficiency shape is evaluated using simulated events as the difference between the results from fitting the generated angular variable distribution of generated events and fitting the reconstructed angular variable distributions of reconstructed events incorporating the efficiency.
Additionally, a systematic uncertainty is estimated to account for possible biases in the fit results when the fit procedure is performed in realistic conditions (presence of background, limited number of available events). A set of 100 independent samples is created with the signal obtained from MC simulation and the background generated from the corresponding \pdf, with signal and background yields matched to the respective values estimated from the data. 
Each of the samples is fit with the standard procedure and the average value over all 100 samples is calculated for each parameter. The systematic uncertainty is obtained by subtracting the average value from the value measured in the fit to the full signal-only MC simulation sample.

To evaluate the uncertainty due to the misidentification fraction being measured using simulated samples, the Gaussian constraint on the parameter $R$ is removed in the fit to the \bzKstarjpsi control sample, yielding an unconstrained measurement of $R$. 
Then, a set of alternative fits is performed, using the measured $R$ values (1.07, 1.02, and 1.02 for 2016, 2017, and 2018, respectively) as the means for the Gaussian constraints in the signal fits.

The systematic uncertainty associated with possible differences in the mass resolution between data and simulation is evaluated by first measuring a resolution scale factor with the \bzKstarjpsi control sample and then applying the same scale factor in the fit to the signal bins. The difference with respect to the nominal result is taken as a systematic uncertainty.

Systematic uncertainties are assessed for the signal and background mass shapes.
To evaluate the uncertainty related to the choice of the signal mass model, a fit is performed to the data using the correctly identified and misidentified mass spectra from MC simulation as templates for the \scpdf and \smpdf components of the \pdf.
To evaluate the uncertainty related to the choice of the background mass model, a fit is performed to the data using Bernstein polynomials. 
For the highest \qsq bin only, the exponential function corrected for the threshold effect at the low edge of the mass range was also used as an alternative function.

The efficiency is evaluated on a finite sample of simulated events. A set of efficiency functions reflecting the statistical variation of the simulated event sample is generated, and used to fit the data. The
standard deviation of the results obtained on each angular parameter is included as a systematic uncertainty. 
As the angular parameters of the background function are fixed in the fit, the statistical uncertainty related to the background parameters is evaluated separately. This is done by varying the parameters using the covariance matrix of the fit to the sidebands to generate 100 different background functions. The uncertainty is obtained by taking the standard deviation of the results obtained by fitting the data with each of the alternative background functions.

The effect of data-to-simulation differences on the angular efficiency and resolution is assessed by performing a fit on the \PJGy control sample and calculating the difference ($\Delta\fl^\PJGy$) between the \fl value obtained from this fit and its current world average~\cite{ParticleDataGroup:2024cfk}. 
This difference is then used to fit the signal sample, with the \fl parameter fixed to $\fl^{\mathrm{nom}} + \Delta\fl^\PJGy$, where $\fl^{\mathrm{nom}}$ is the value of \fl obtained from the nominal fit to the signal sample. The value of $\Delta\fl^\PJGy$ and the deviation of the other fit parameters from their nominal values are assigned as systematic uncertainties.

The bias introduced by the contamination from partially reconstructed background sources in the lower-mass sideband on the background angular shape is evaluated in the \PJGy control sample, excluding from the mass sidebands events with the \PBz candidate mass lower than 5.1\GeV to define the background angular shape.
The difference between the new results and the nominal ones is considered as a systematic uncertainty.
Finally, the uncertainty due to the possible contamination of events from the two resonant channels in the adjoining \qsq bins is evaluated by adding a component with a simulation-based shape to the \pdf to describe the contribution of the resonant channels, and using the difference of the results as the systematic uncertainty. 

The systematic uncertainties are determined for each \qsq bin, and in cases where the systematic uncertainty is obtained as the difference between a nominal and alternative configuration, the systematic uncertainty is symmetrized. Their impact on the various angular parameters is summarized in Table~\ref{tab:syst_unc}.
The individual systematic uncertainties are considered independent and added in quadrature.

\begin{table*}[htp!]
  \topcaption{The uncertainties considered in the analysis on the various angular observables. 
  For each source of uncertainty, the range covers the absolute variation observed across the \qsq bins.
  \label{tab:syst_unc}}
  \centering
    \begin{tabular}{l c c c c c c c c}
      Source & \fl ($\times 10^{-3}$) & \pone ($\times 10^{-3}$) & \ptwo ($\times 10^{-3}$) & \pthree ($\times 10^{-3}$) \\\hline
Efficiency modeling        & 1--9     & 7--44  & 3--11    & 0--46  \\ 
Fit bias                   & 1--2     & 0--6   & 2--62    & 1--12  \\ 
Misidentification fraction & 0--2     & 1--4   & 1--3     & 0--14  \\ 
Signal mass resolution     & 1--10    & 1--12  & 2--11    & 1--21  \\ 
Signal mass shape          & 0--9     & 1--22  & 0--10    & 3--70  \\ 
Background mass shape      & 0--5     & 1--16  & 1--13    & 0--8   \\ 
Efficiency (statistical)   & 1--10    & 5--31  & 1--64    & 4--45  \\ 
Background (statistical)   & 2--6     & 4--20  & 1--21    & 2--16  \\ 
Data/simulation differences& 8        & 0--23  & 0--16    & 0--13  \\ 
Partially reco background  & 1        & 1      & 0        & 1      \\ 
Resonant background        & 0--1     & 0--6   & 0--5     & 0--2   \\ 
    \end{tabular}

 \bigskip
     \begin{tabular}{l c c c c c c c c}
      Source      & \pfourp ($\times 10^{-3}$) & \pfivep ($\times 10^{-3}$) & \psixp ($\times 10^{-3}$) & \peightp ($\times 10^{-3}$) \\ \hline
Efficiency modeling        & 3--87   & 2--13   & 5--16  & 6--28  \\
Fit bias                   & 9--54   & 0--8    & 0--3   & 0--24  \\
Misidentification fraction & 1--5    & 1--10   & 0--4   & 0--12  \\
Signal mass resolution     & 4--23   & 0--12   & 0--5   & 0--16  \\
Signal mass shape          & 2--16   & 1--15   & 0--7   & 0--91  \\
Background mass shape      & 6--30   & 1--13   & 0--7   & 1--10  \\
Efficiency (statistical)   & 5--47   & 4--22   & 4--13  & 10--59 \\ 
Background (statistical)   & 6--37   & 4--24   & 3--9   & 5--23  \\
Data/simulation differences& 0--11   & 0--13   & 0--3   & 0--30  \\
Partially reco background  & 25      & 0       & 0      & 2      \\
Resonant background        & 0--30   & 0--11   & 0--5   & 0--12  \\
    \end{tabular}
\end{table*}

\section{Results}
\begin{figure*}[hbtp]
  \centering
    \includegraphics[width=\figfourwidth]{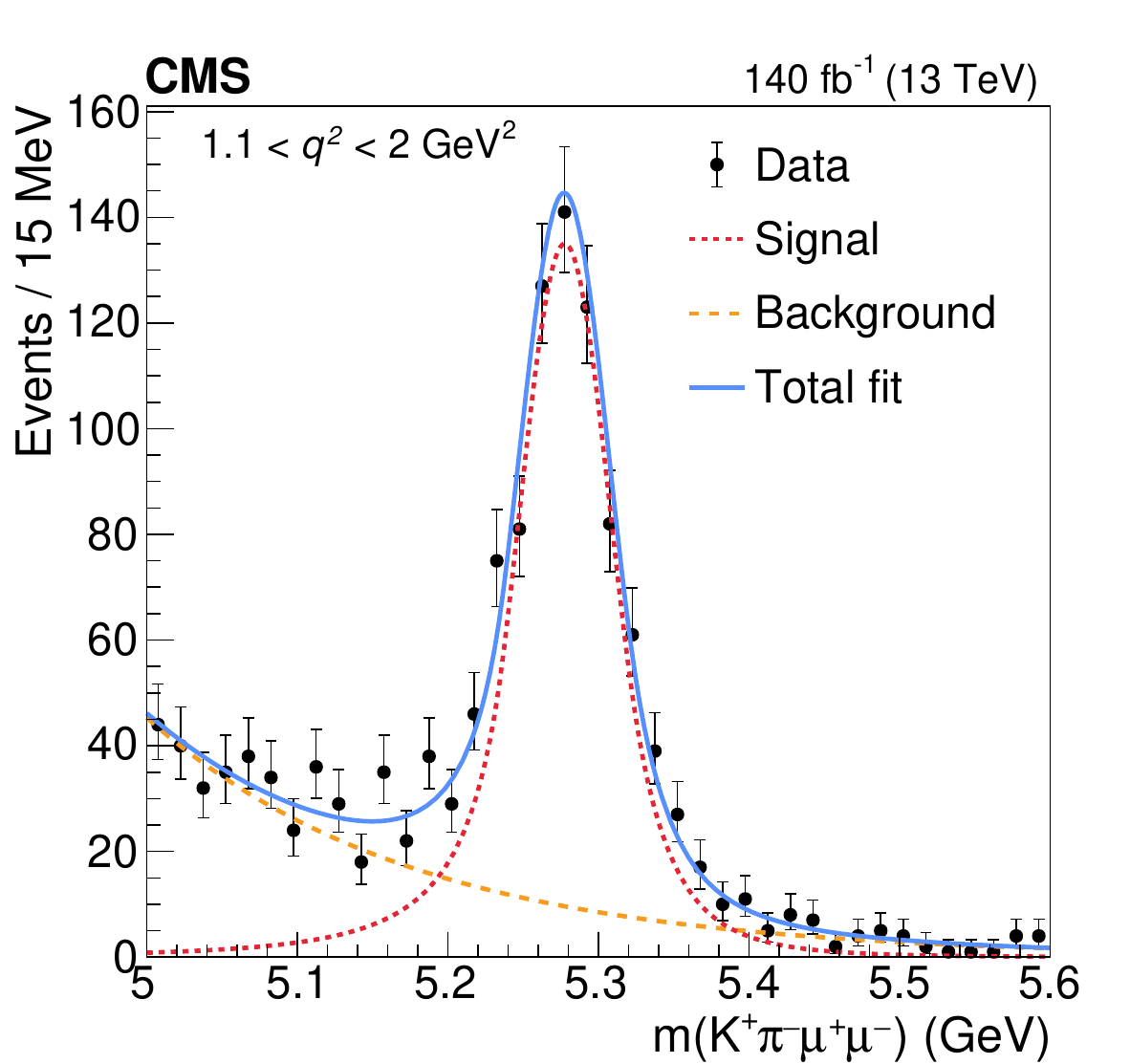}
    \includegraphics[width=\figfourwidth]{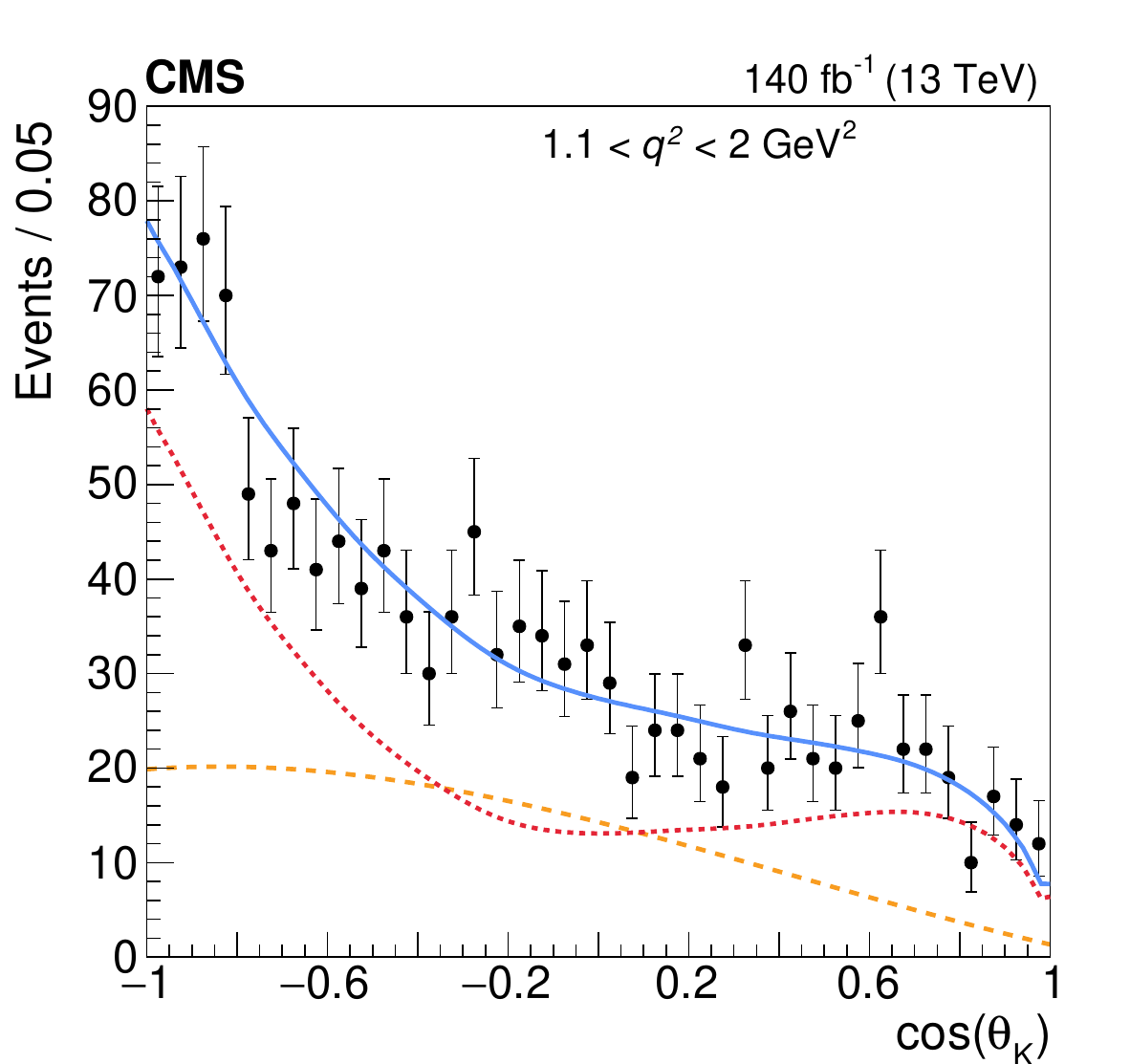}
    \includegraphics[width=\figfourwidth]{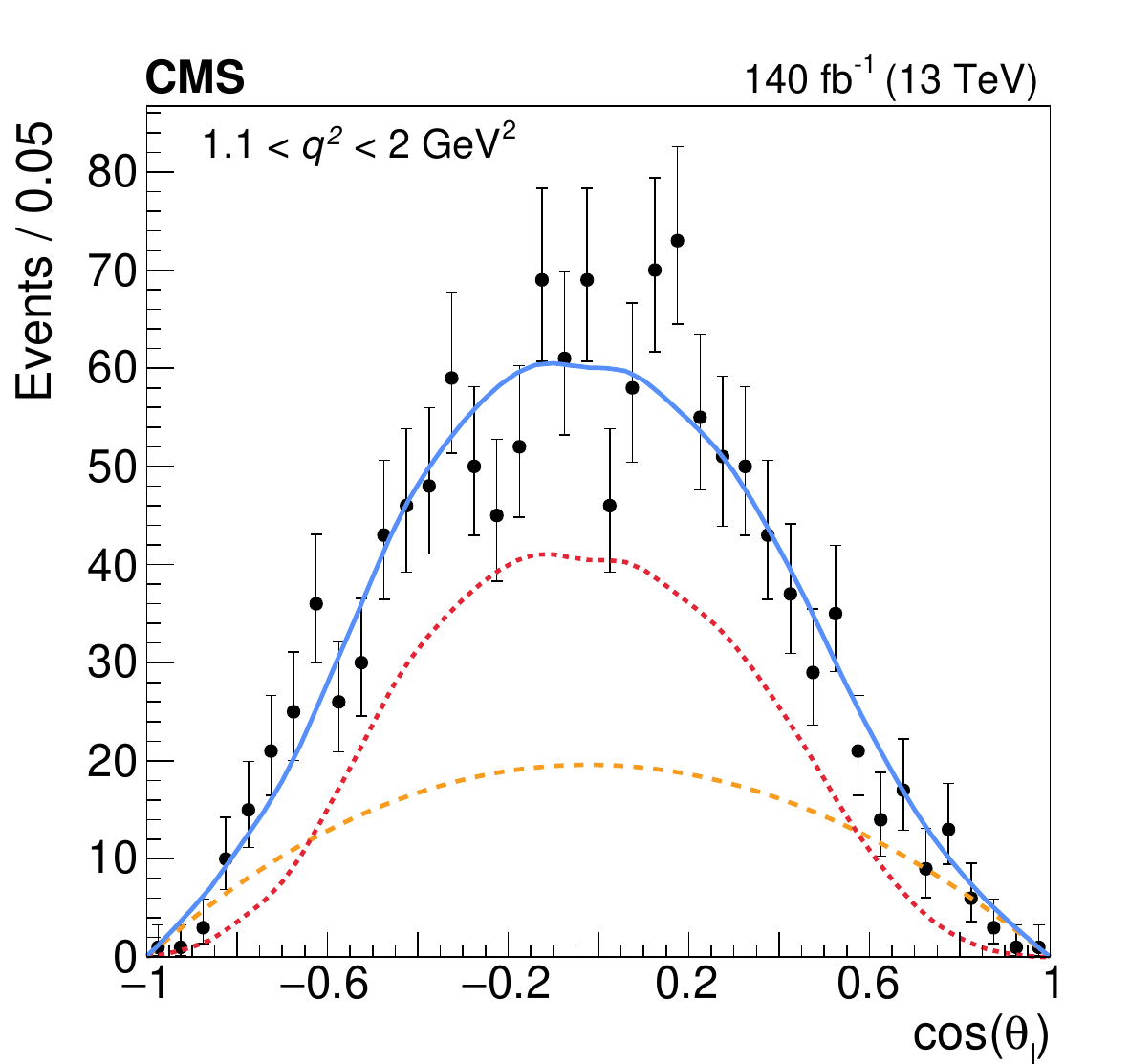}
    \includegraphics[width=\figfourwidth]{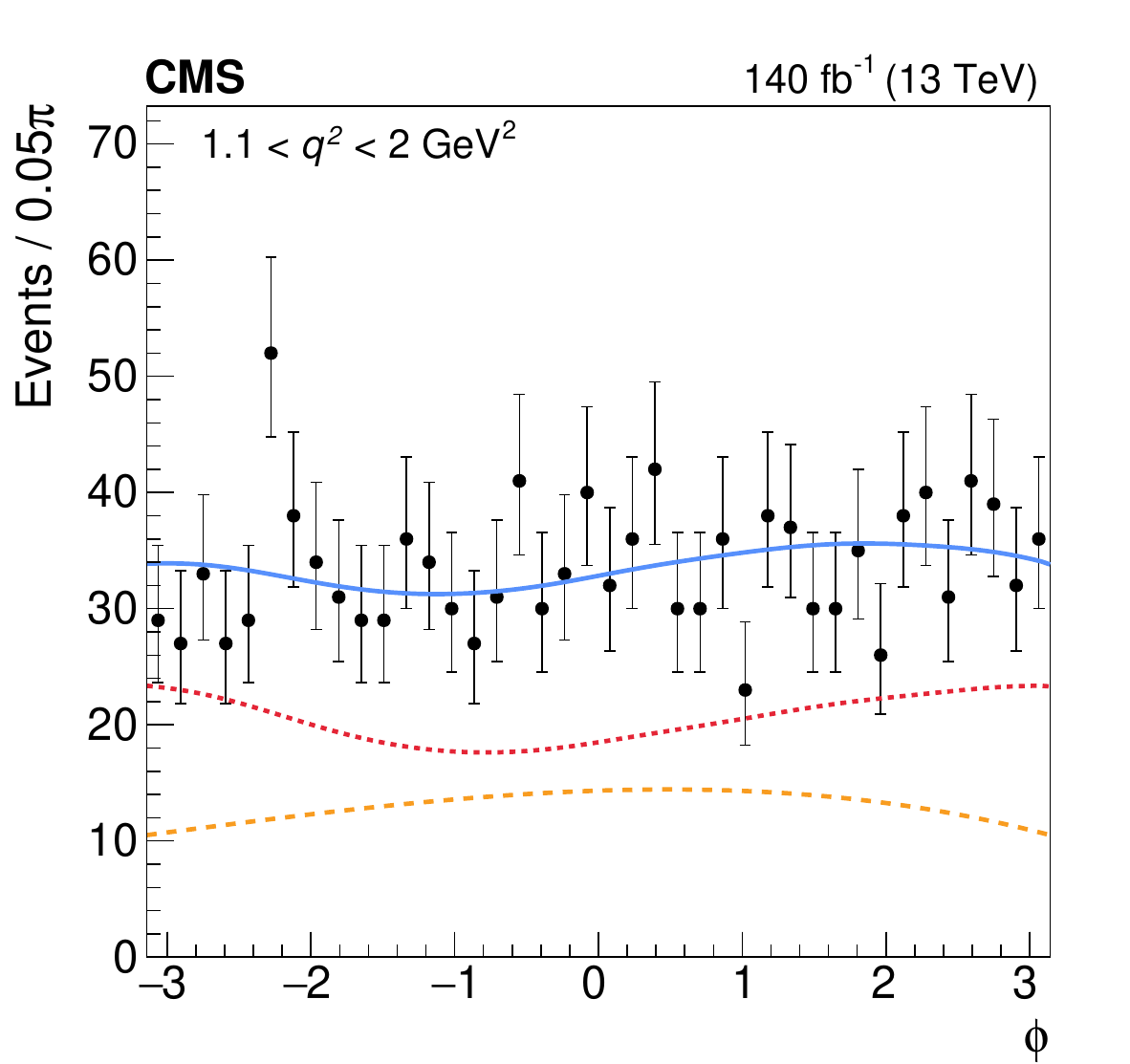}\\
    \includegraphics[width=\figfourwidth]{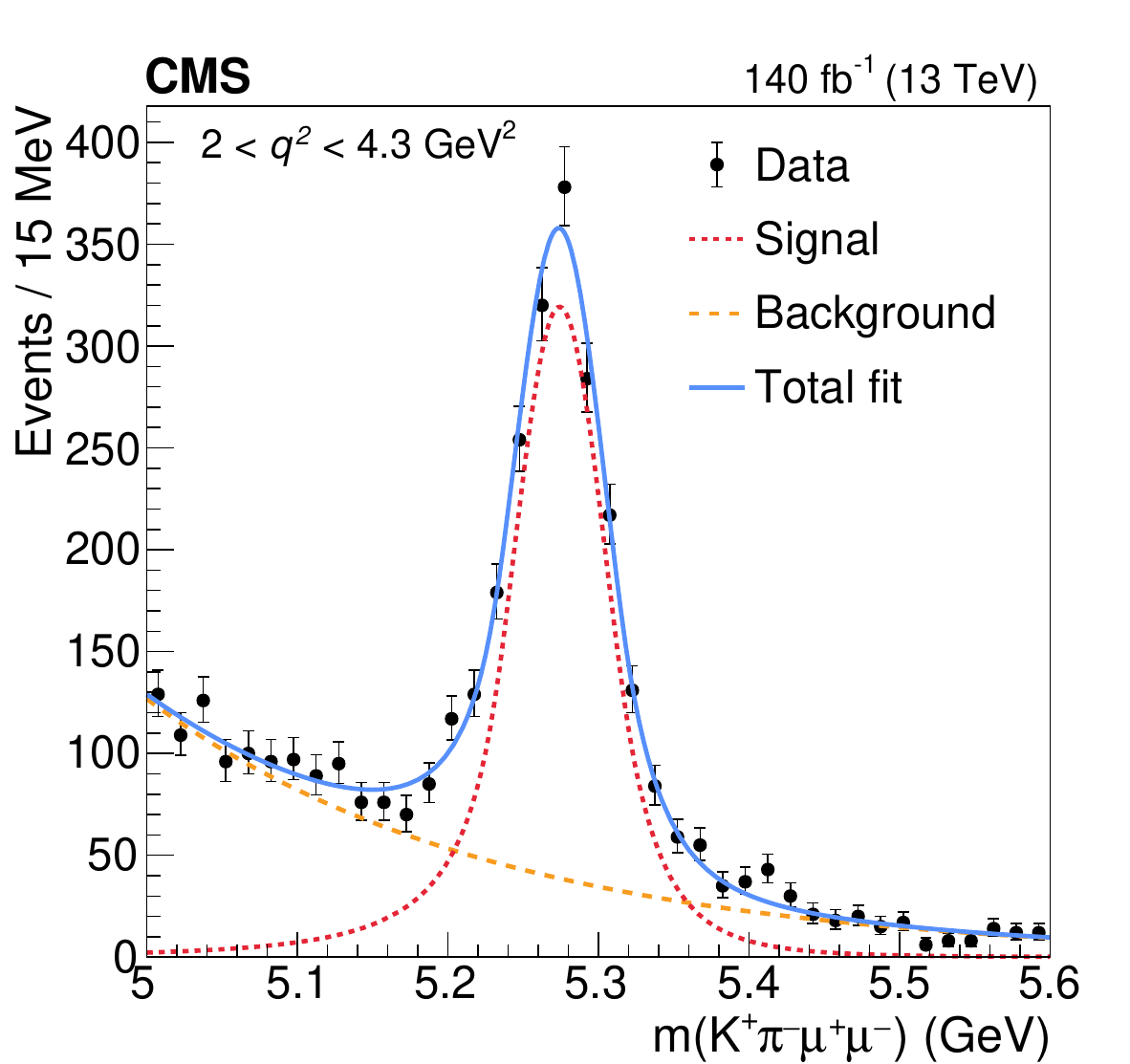}
    \includegraphics[width=\figfourwidth]{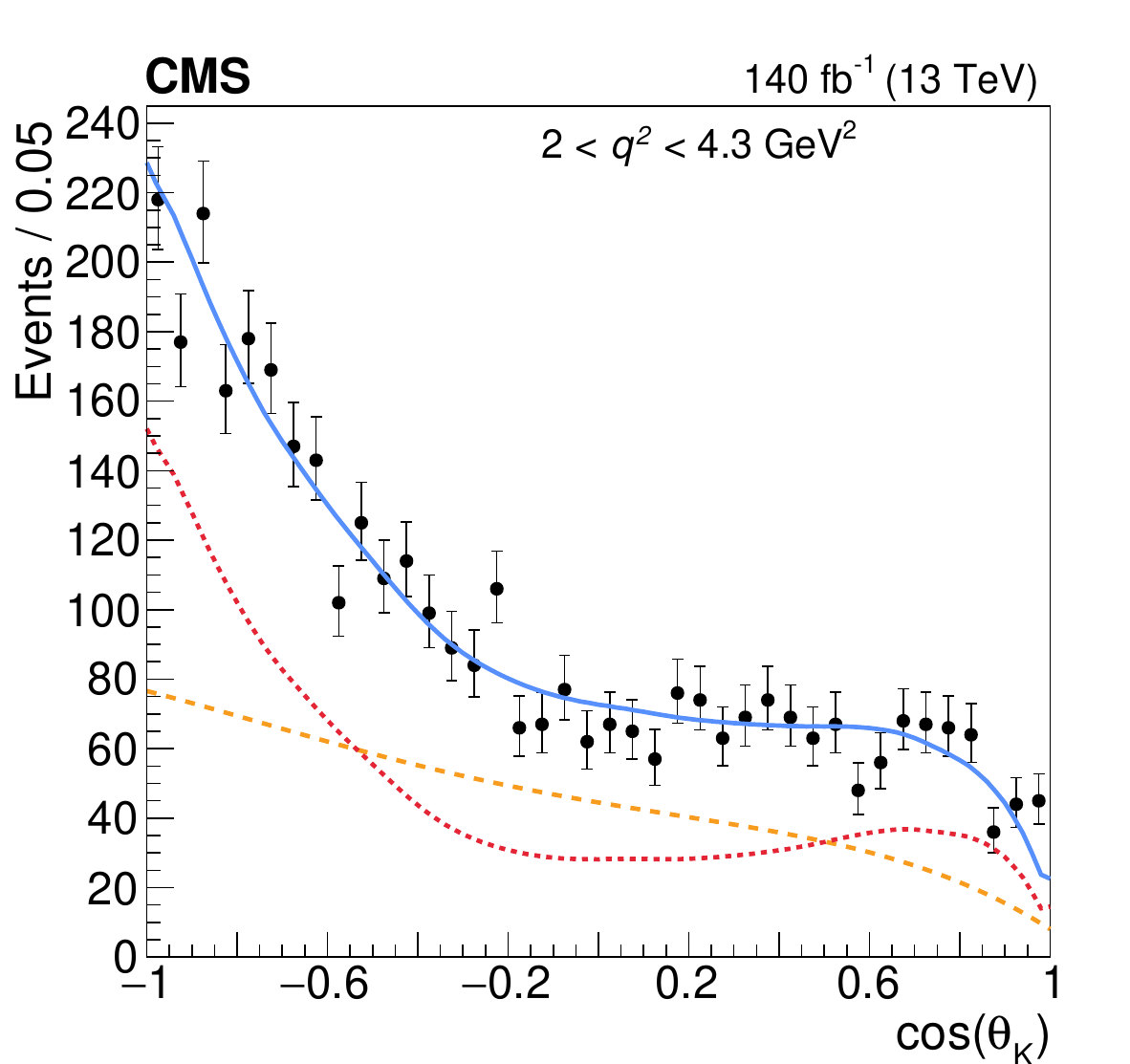}
    \includegraphics[width=\figfourwidth]{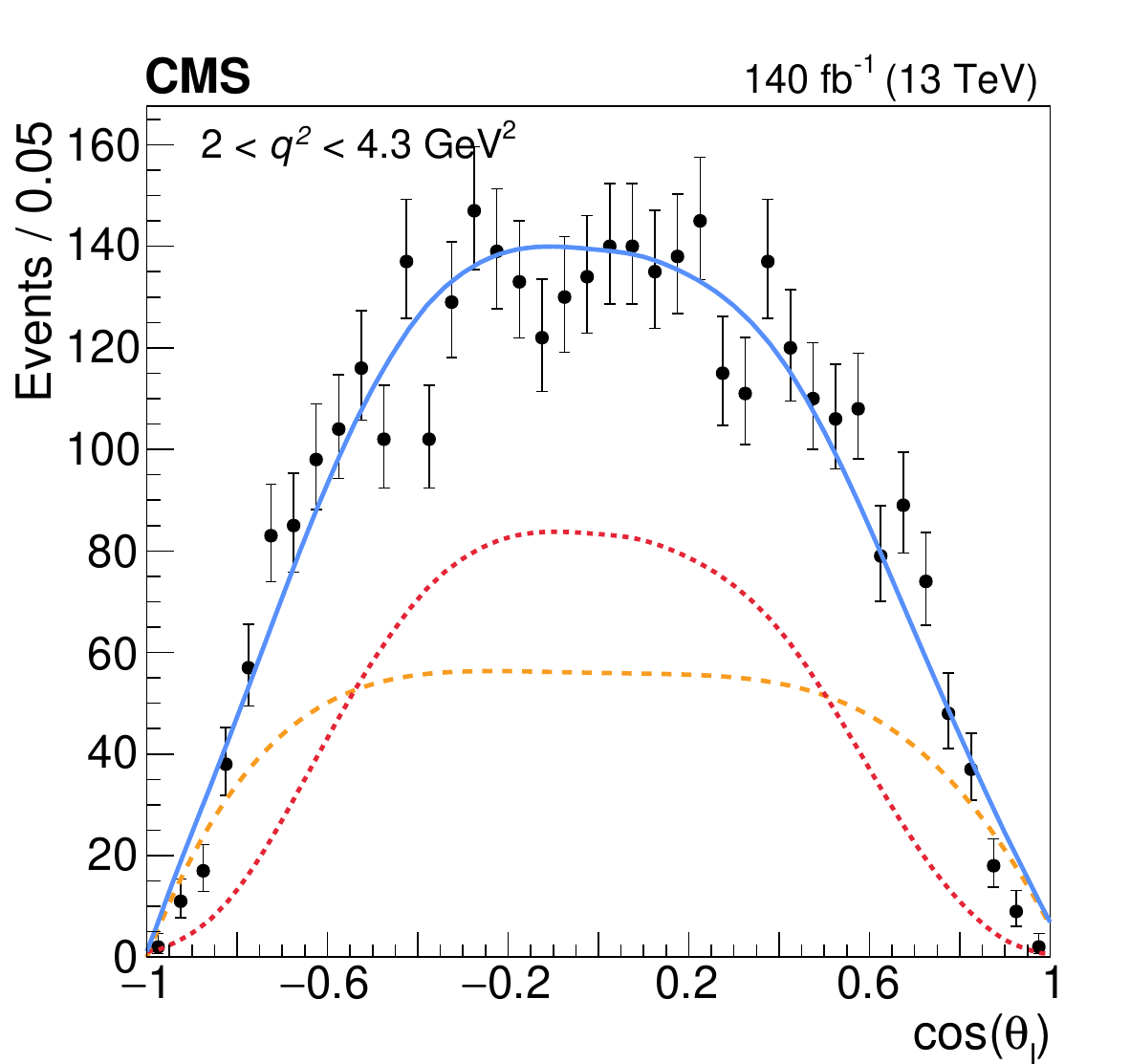}
    \includegraphics[width=\figfourwidth]{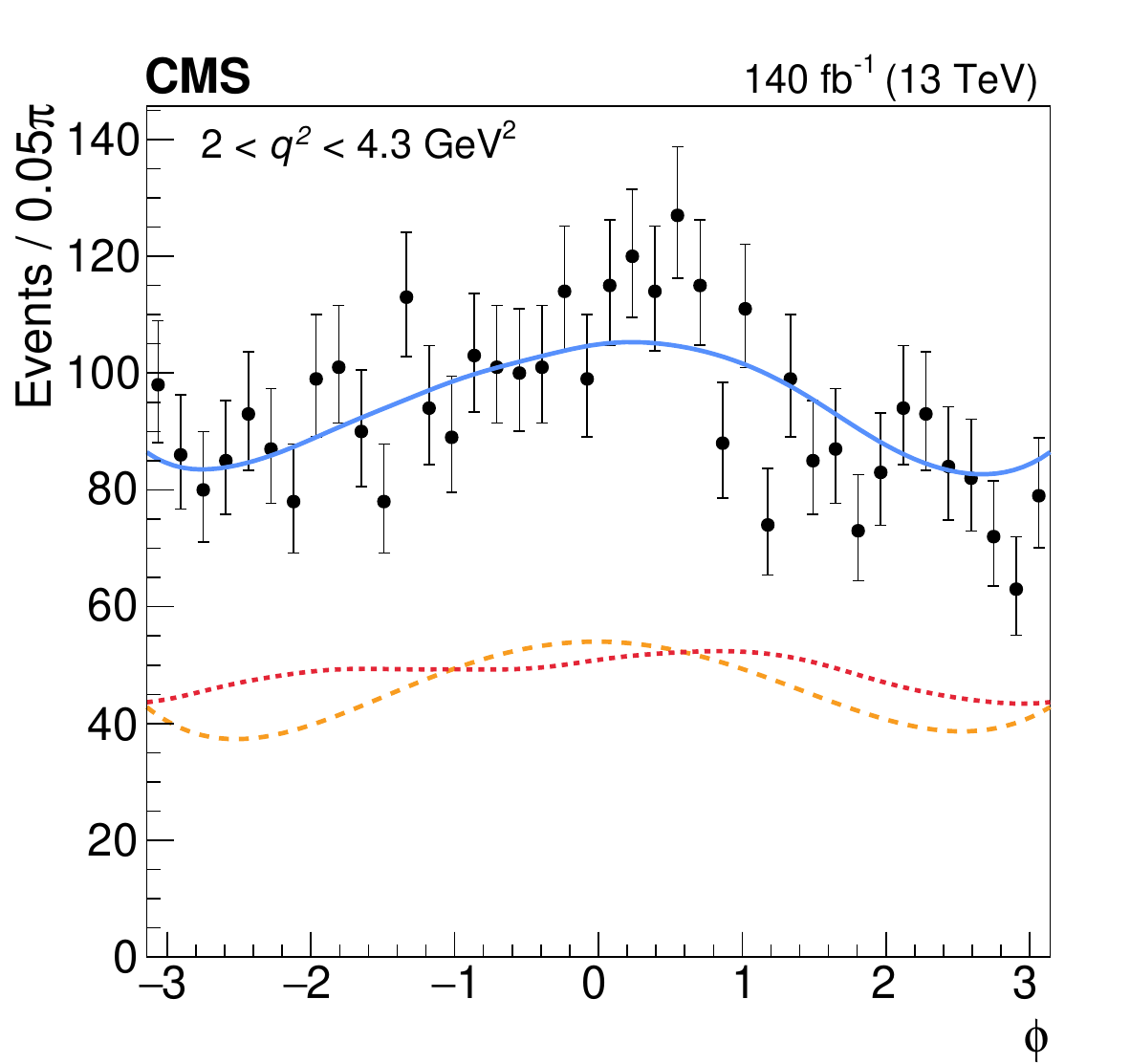}\\
    \includegraphics[width=\figfourwidth]{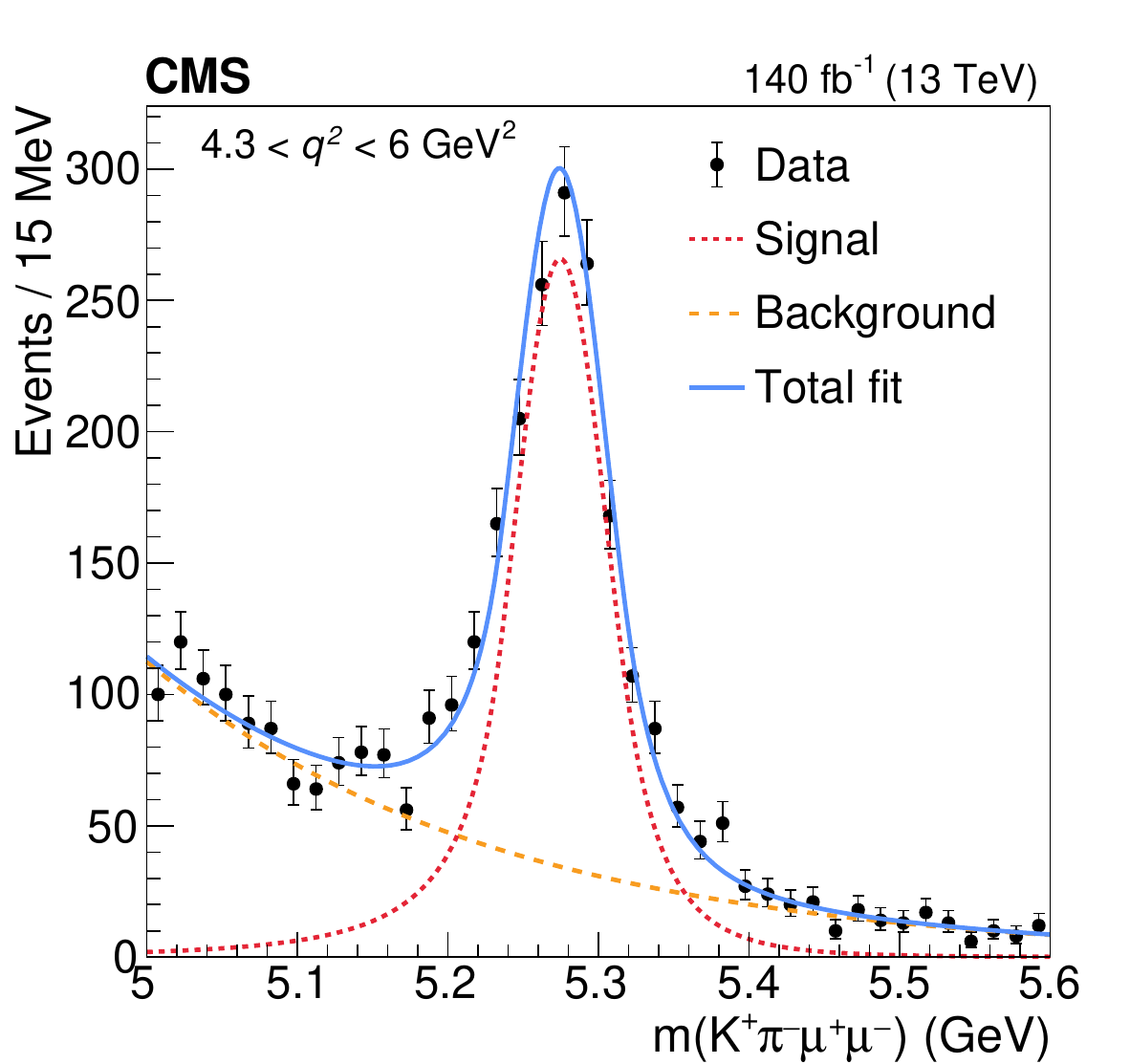}
    \includegraphics[width=\figfourwidth]{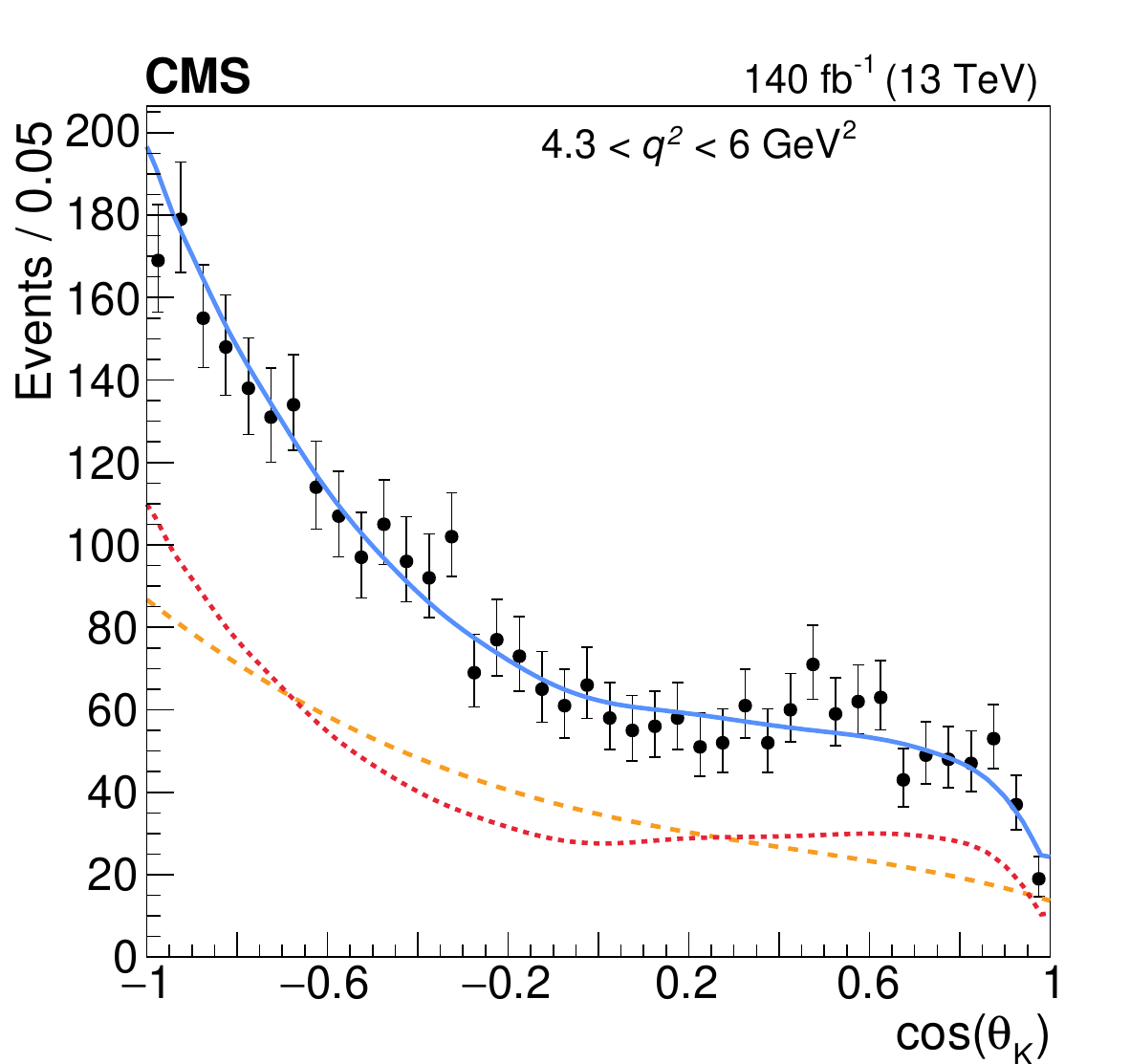}
    \includegraphics[width=\figfourwidth]{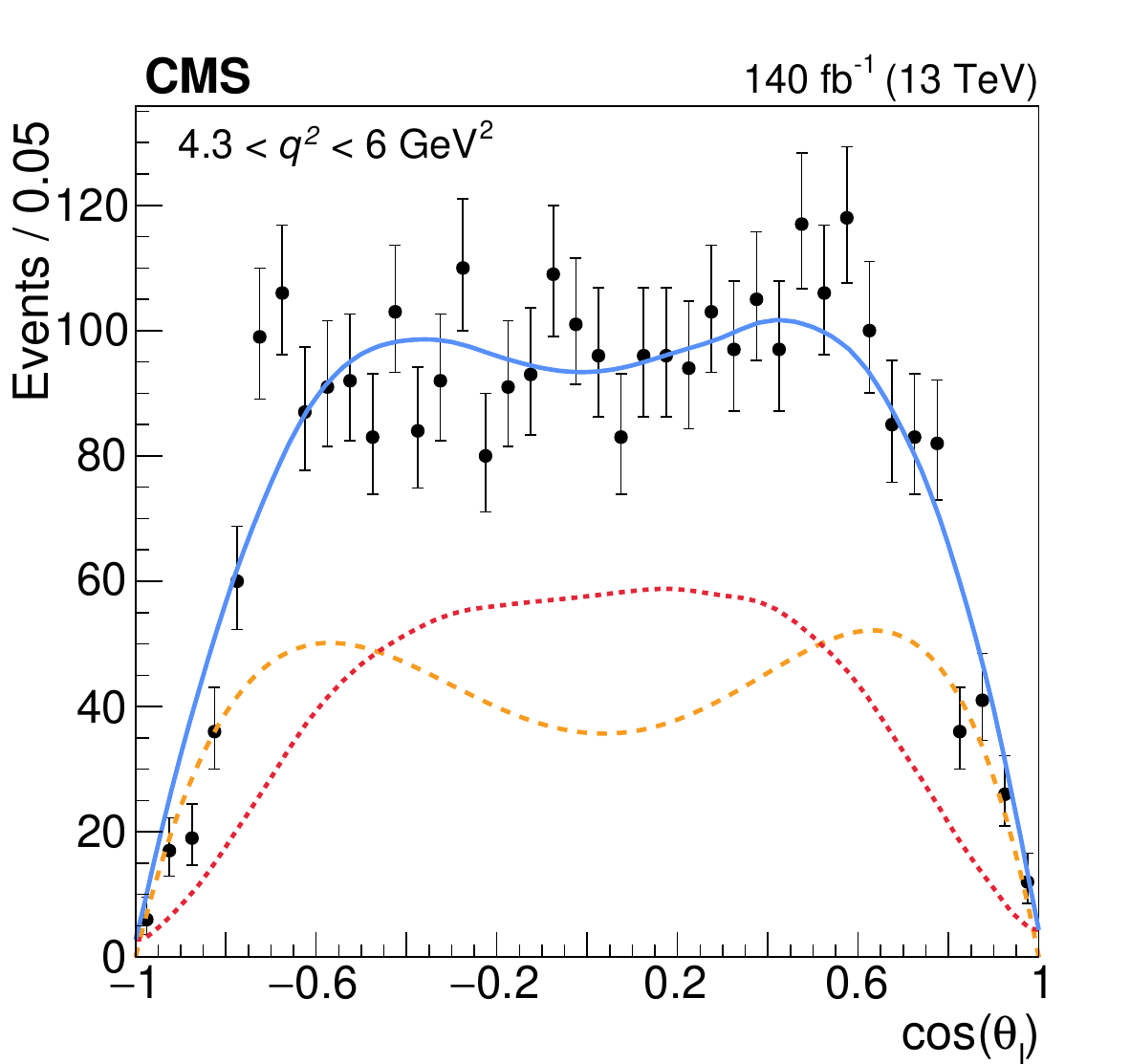}
    \includegraphics[width=\figfourwidth]{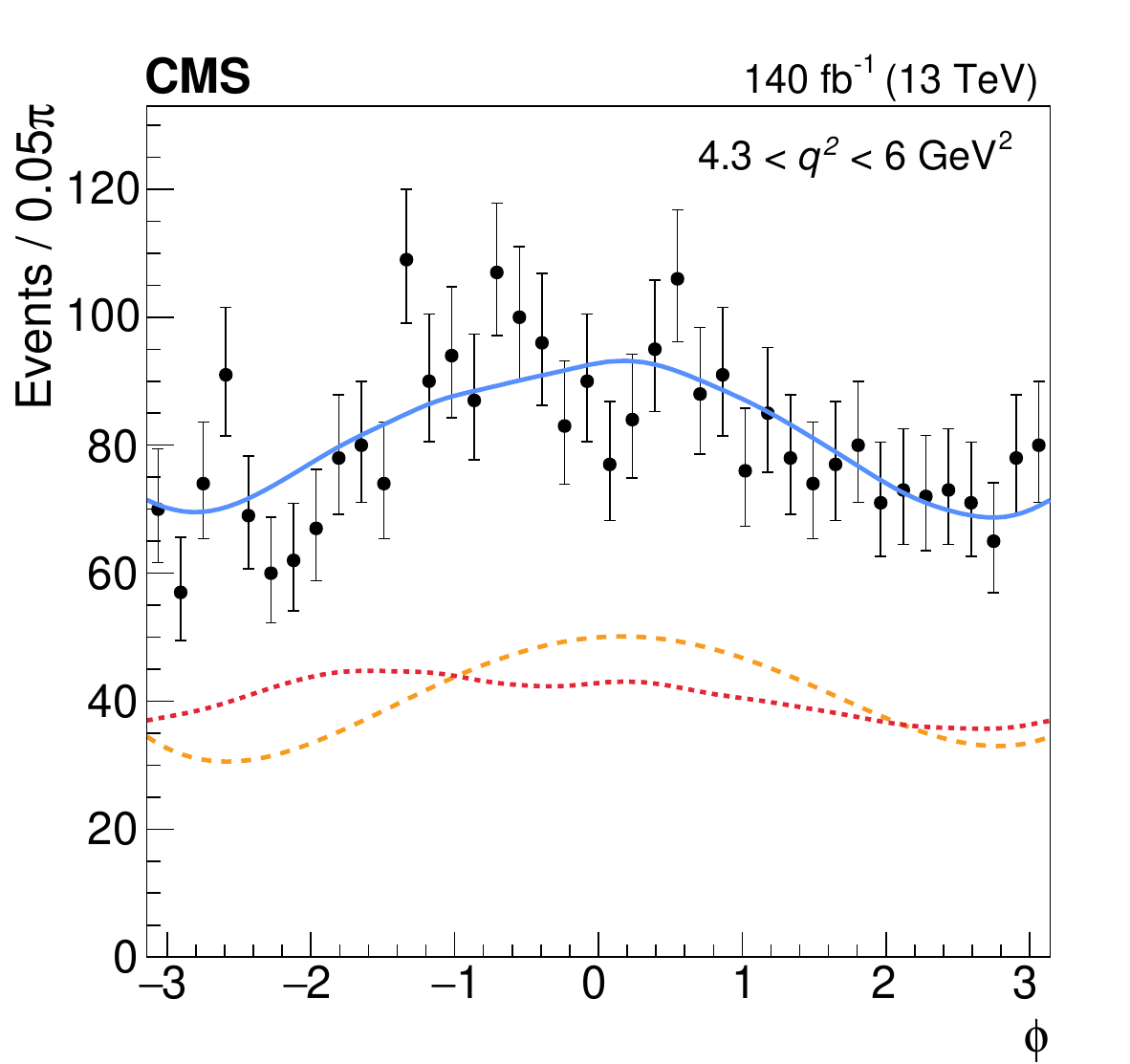}\\
    \includegraphics[width=\figfourwidth]{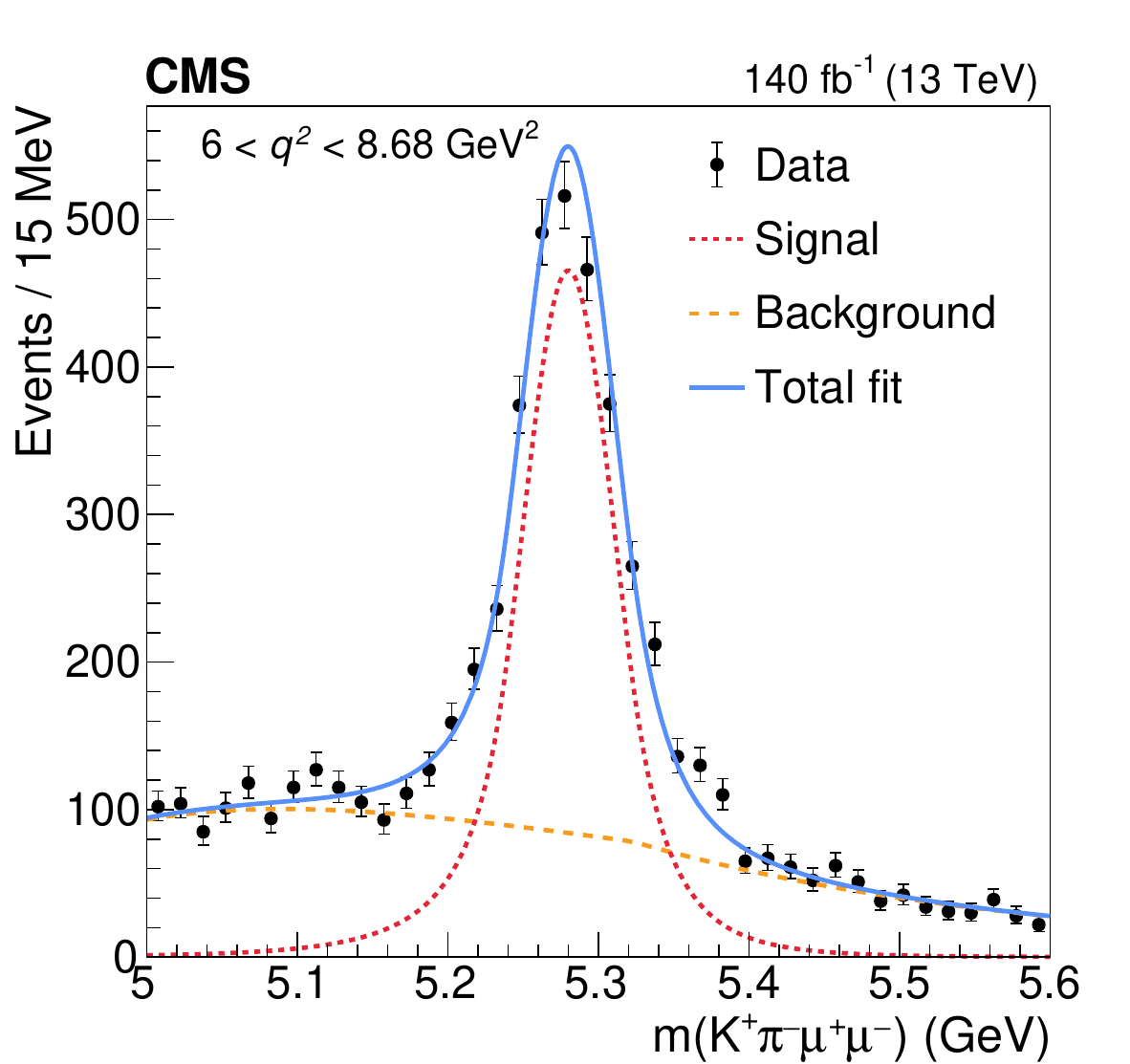}
    \includegraphics[width=\figfourwidth]{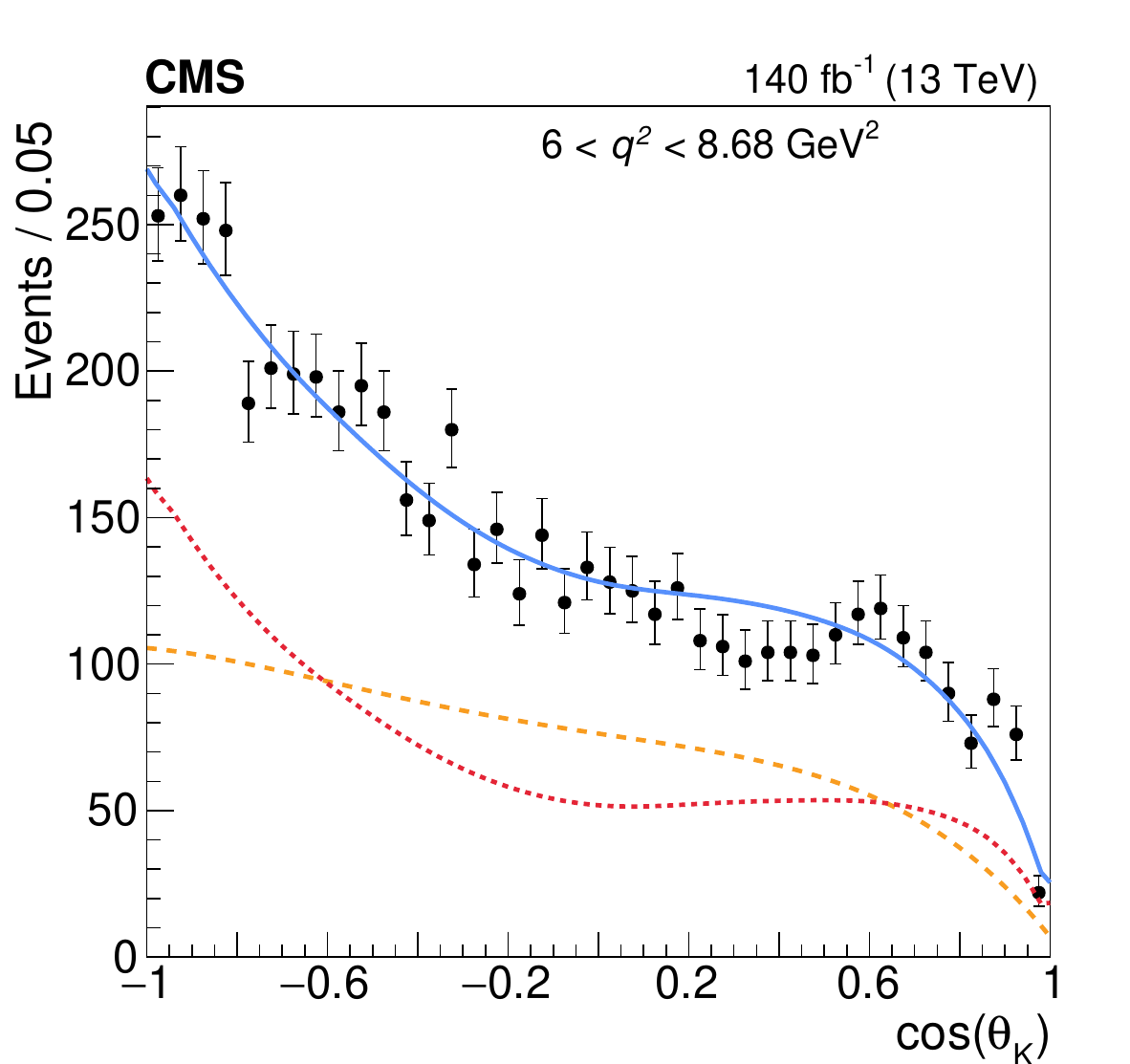}
    \includegraphics[width=\figfourwidth]{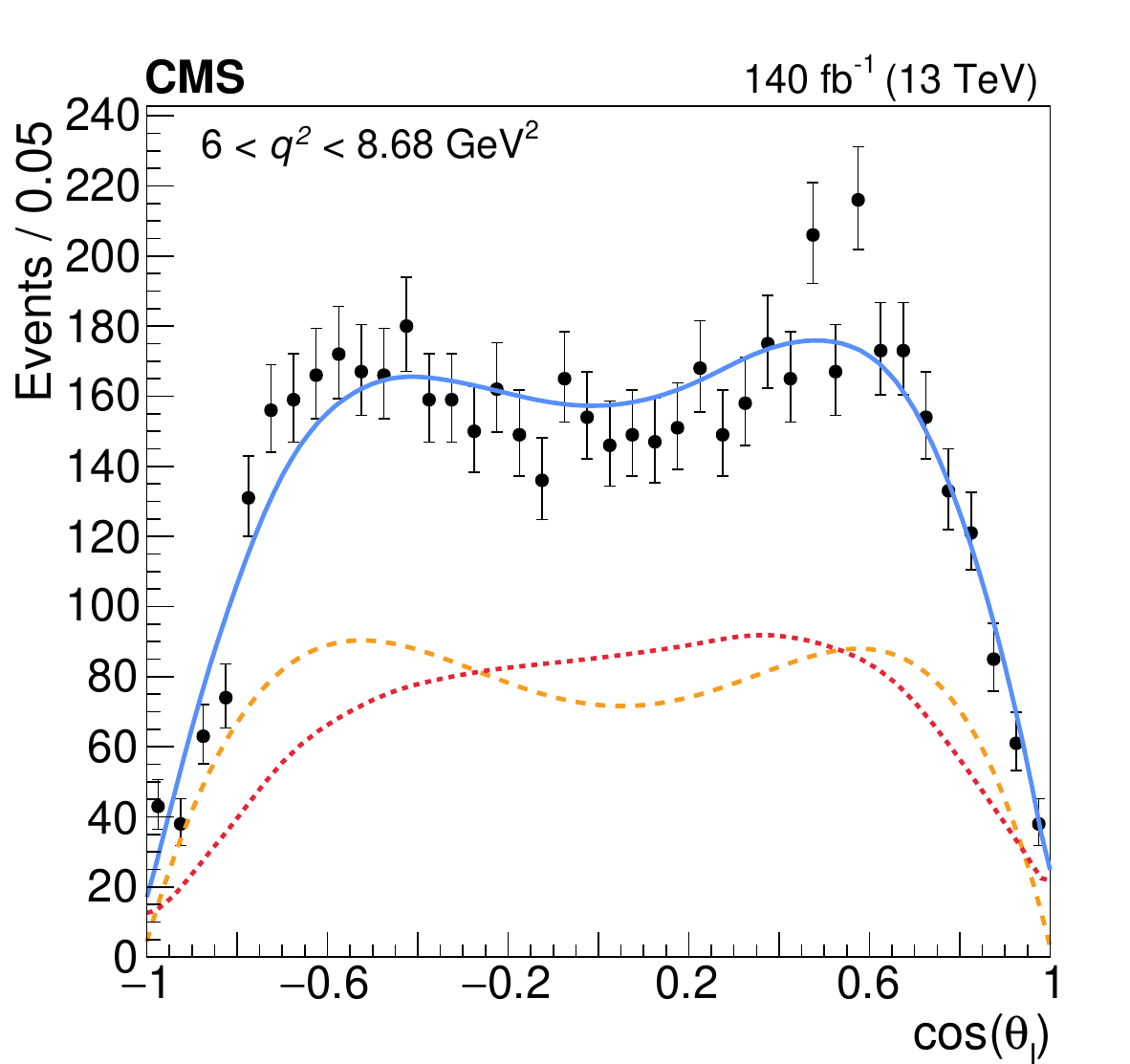}
    \includegraphics[width=\figfourwidth]{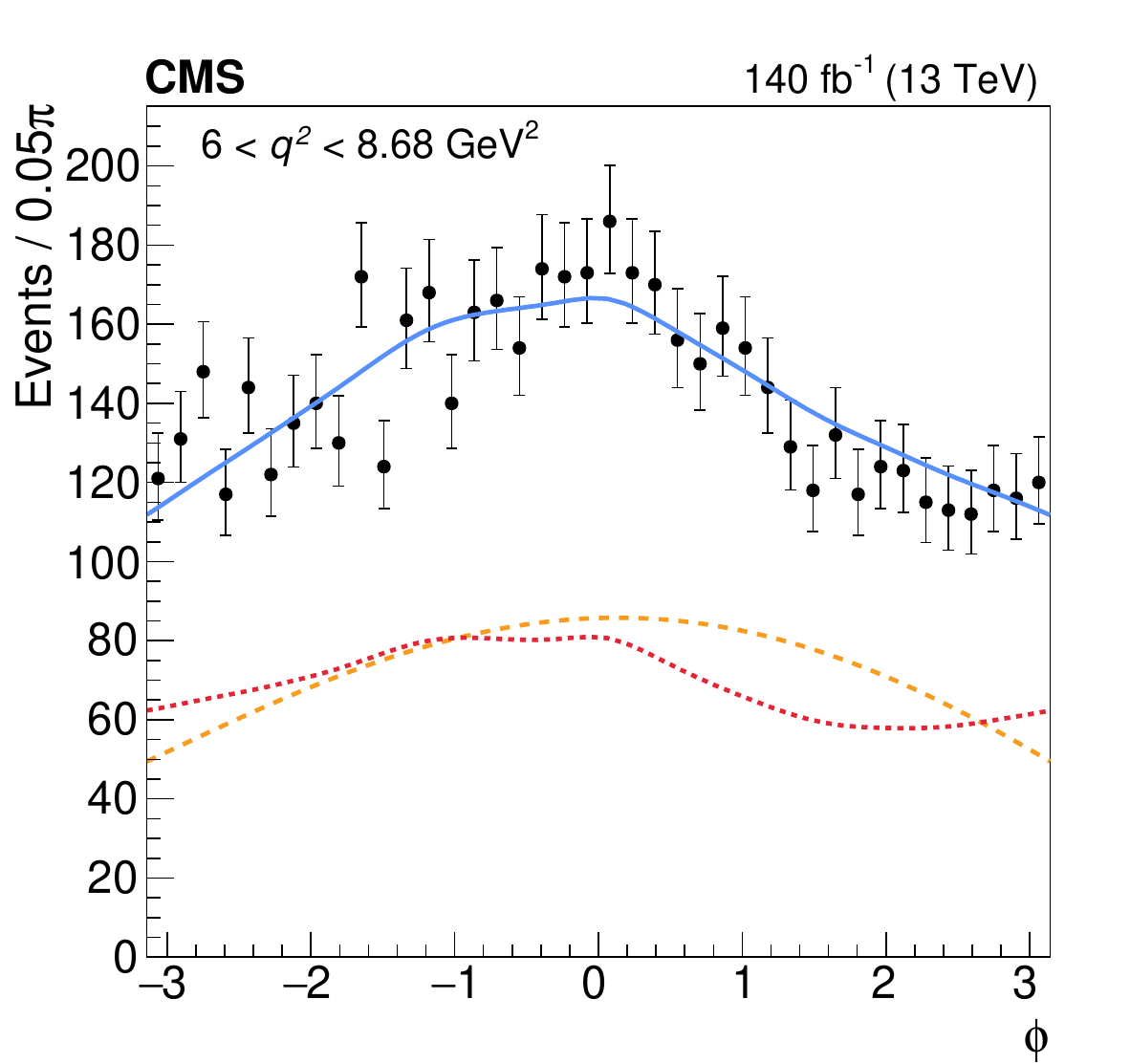}\\
    \includegraphics[width=\figfourwidth]{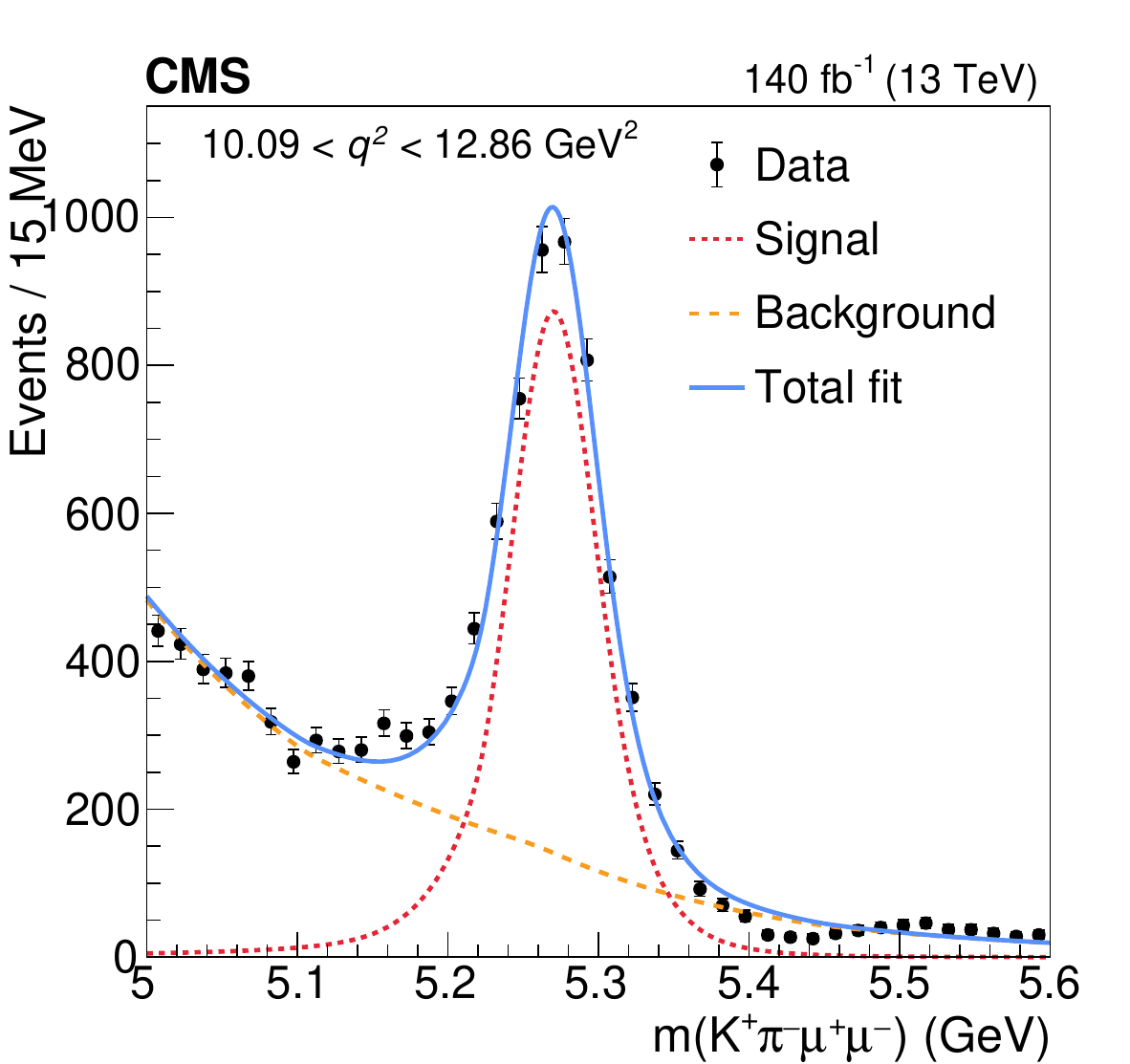}
    \includegraphics[width=\figfourwidth]{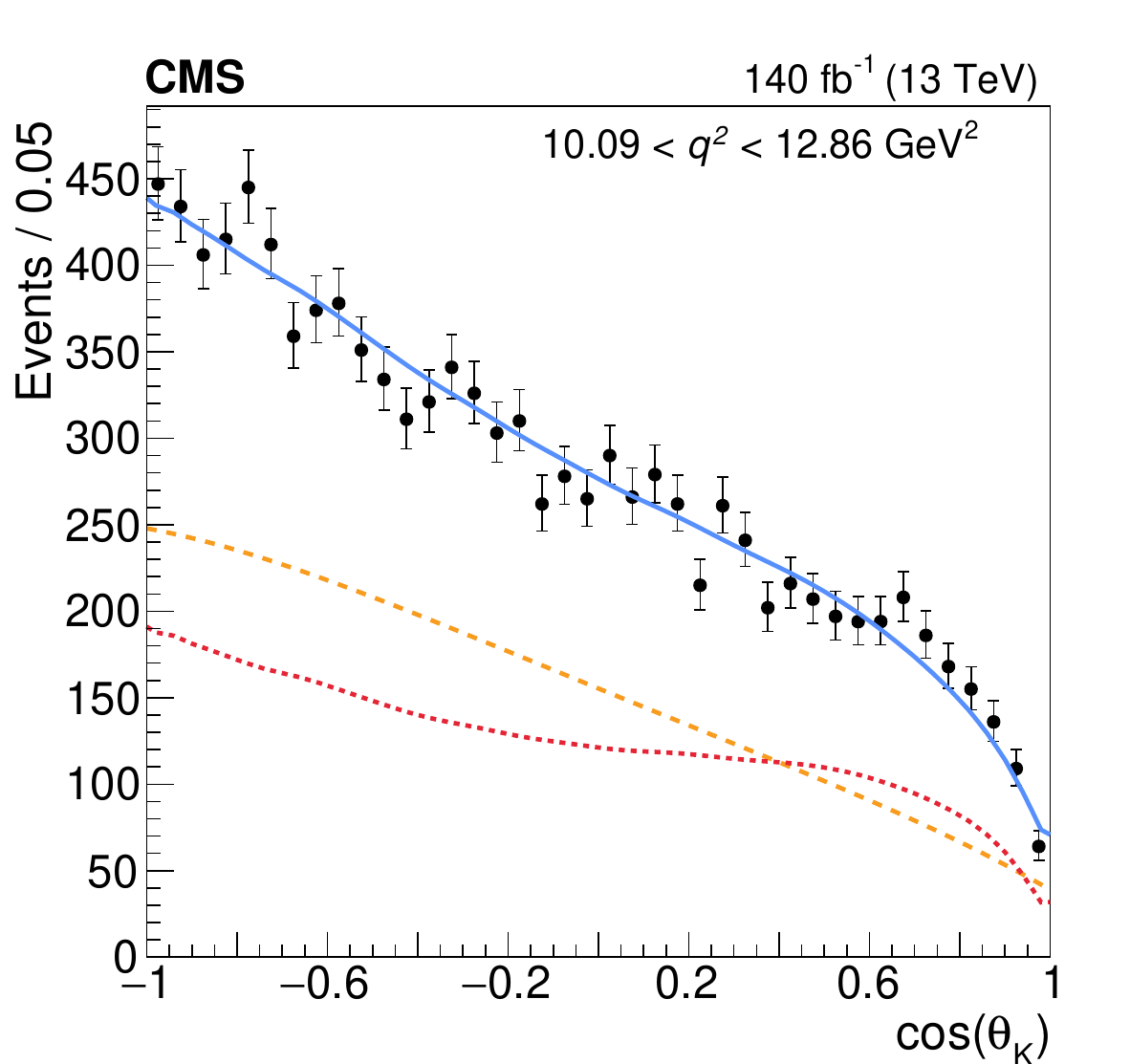}
    \includegraphics[width=\figfourwidth]{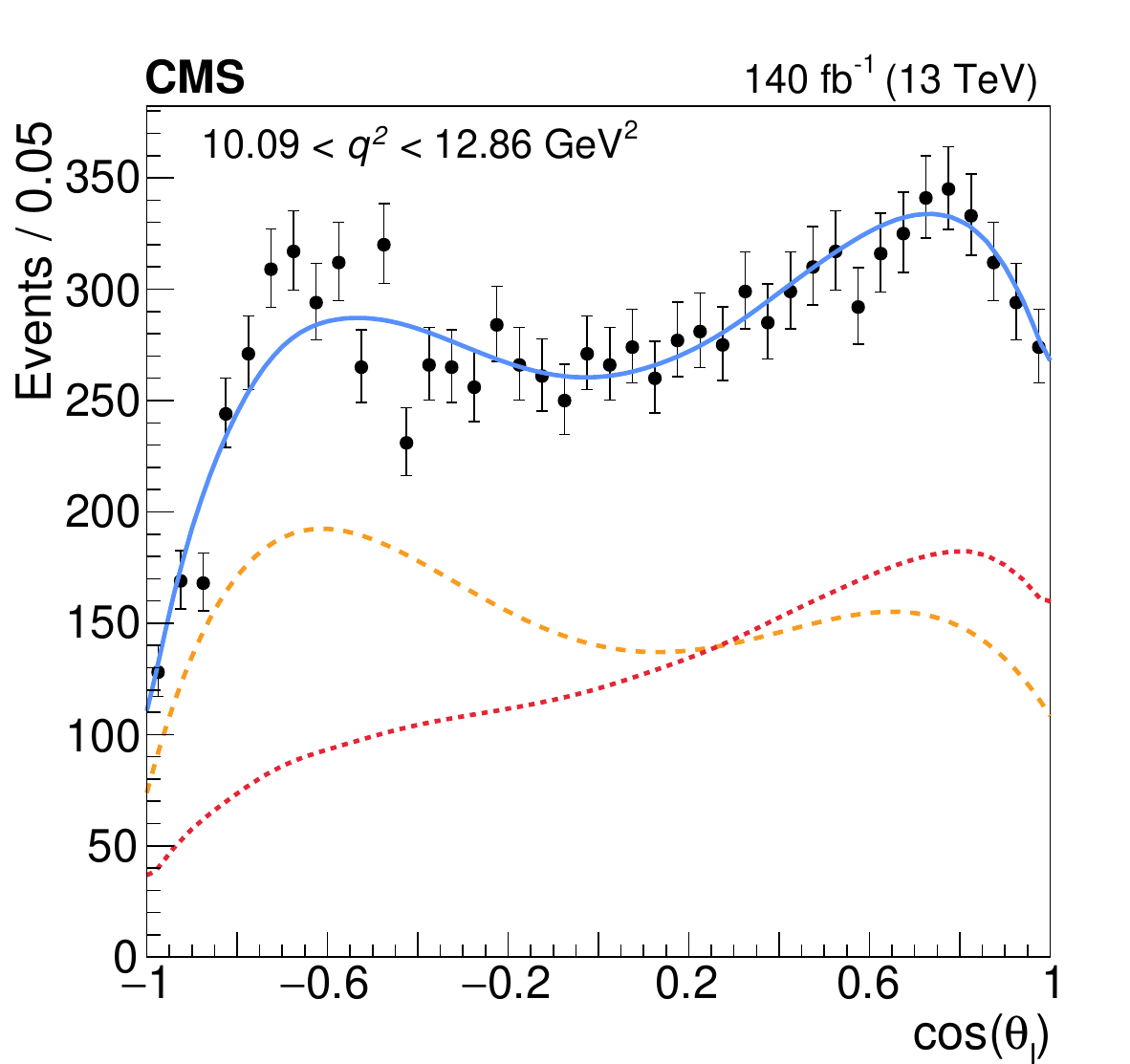}
    \includegraphics[width=\figfourwidth]{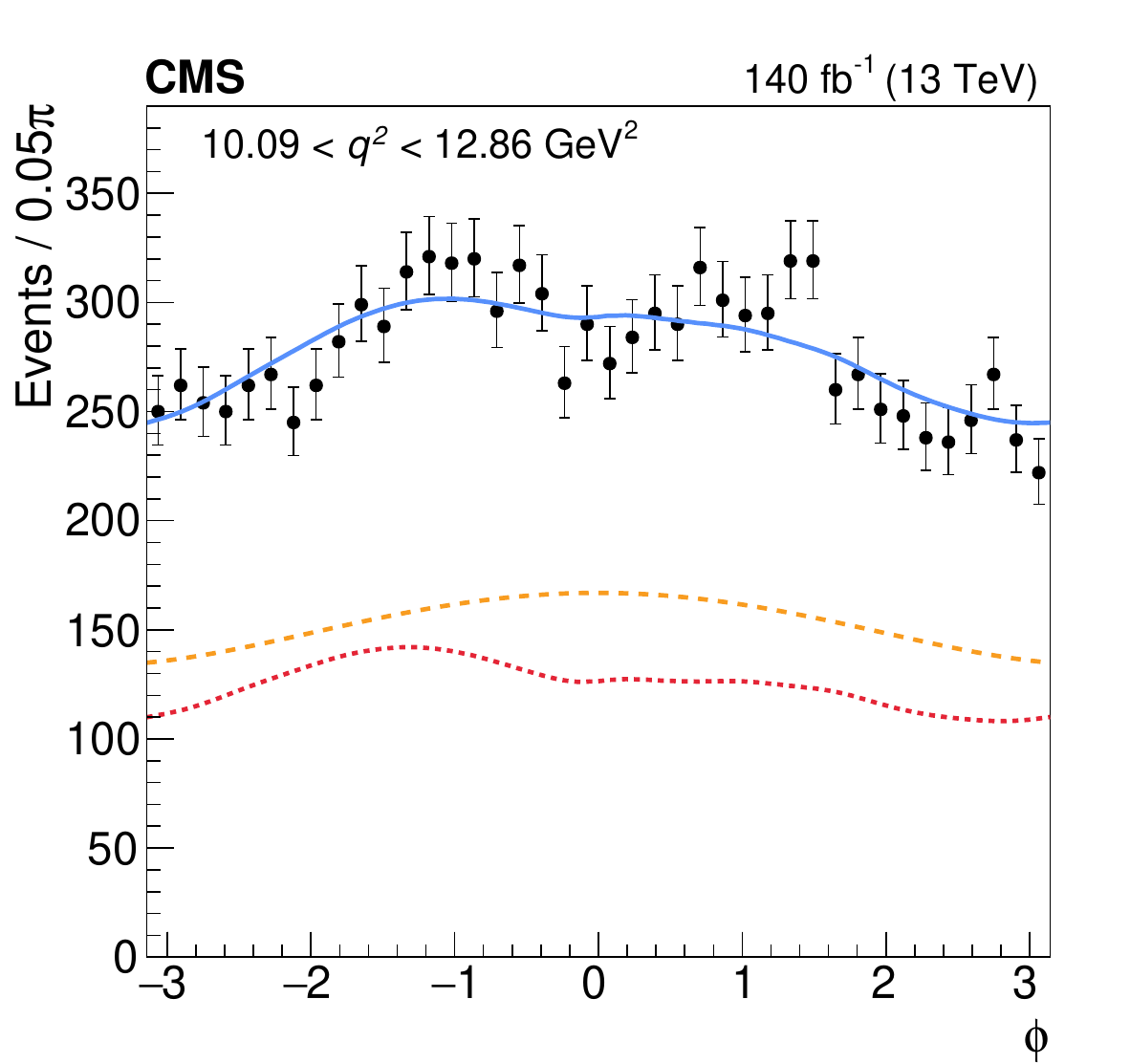}\\
    \includegraphics[width=\figfourwidth]{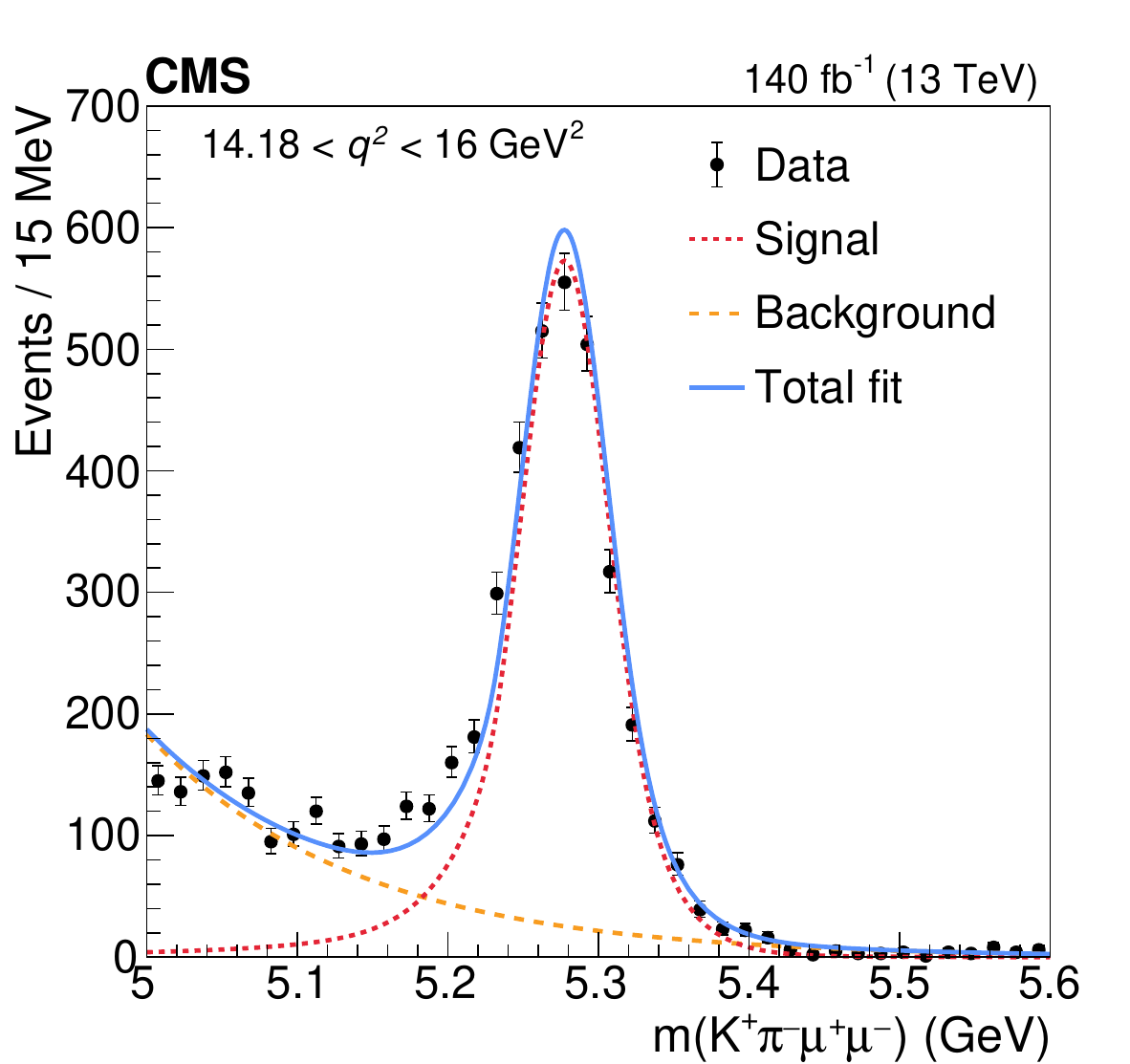}
    \includegraphics[width=\figfourwidth]{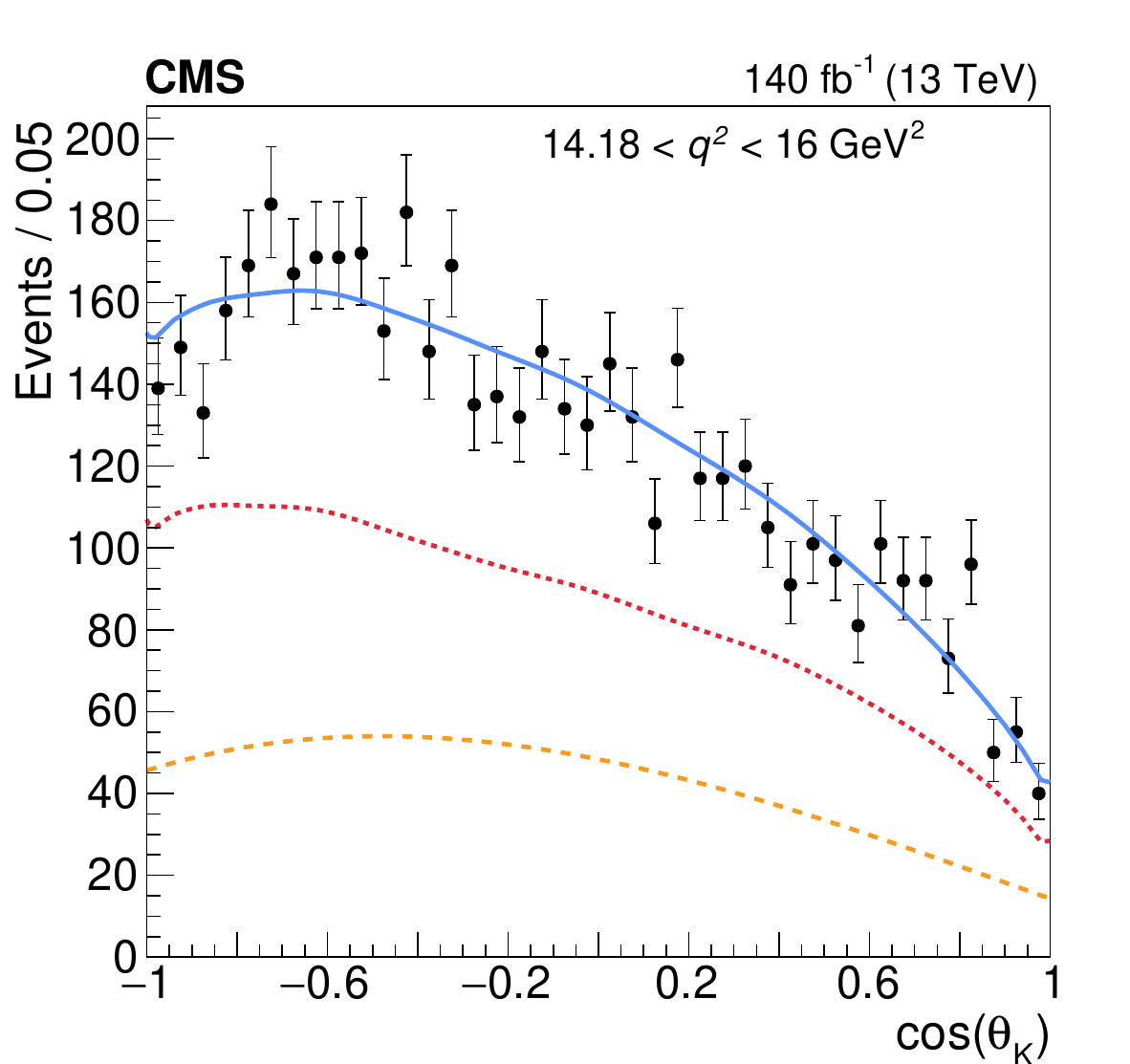}
    \includegraphics[width=\figfourwidth]{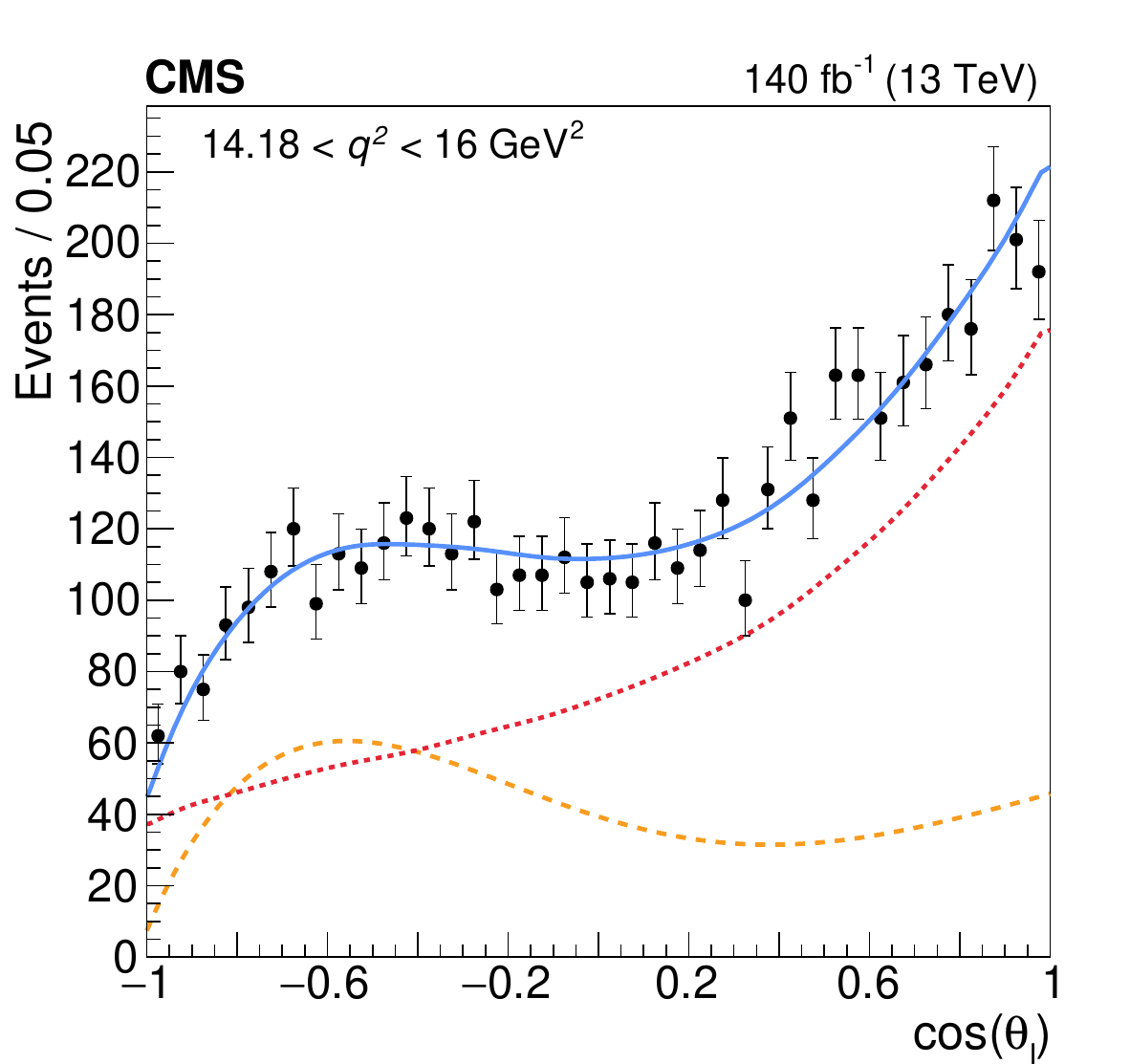}
    \includegraphics[width=\figfourwidth]{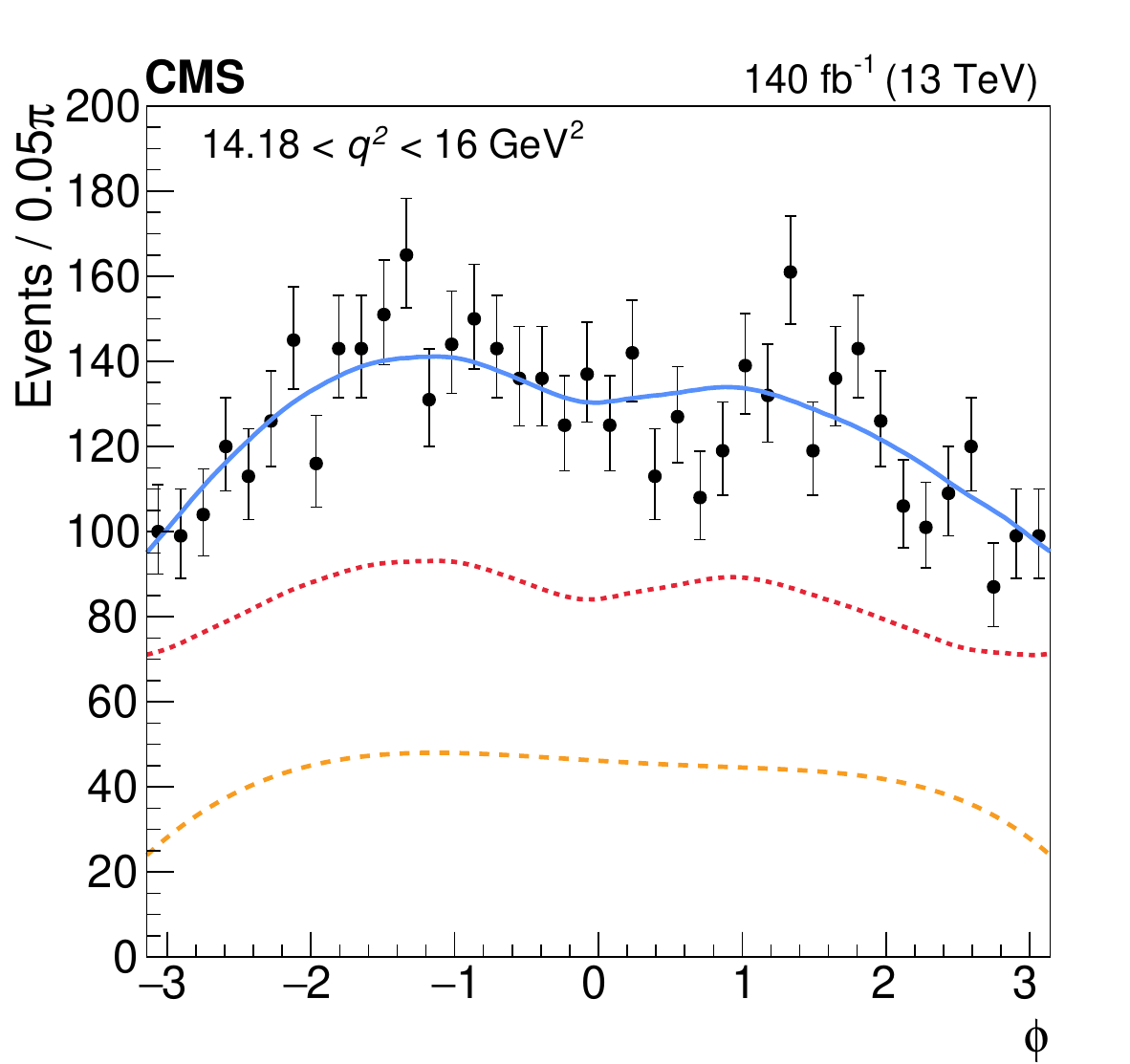}\\
    \caption{Mass and angular distributions for the six \qsq bins (one bin per row). 
    The projections of the total fitted distribution (blue solid line) and its different components are overlaid. The signal is shown by the red dotted line, and the background by the orange dashed line.}
    \label{fig:proj_bins}
\end{figure*}

The projections of the fitted \pdf on the mass and angular distributions are shown in Fig.~\ref{fig:proj_bins} for all \qsq bins.
The fit quality has been verified in all \qsq bins by assessing the agreement between the \pdf projections and the data. 

\begin{figure*}[hbtp!]
  \centering
    \includegraphics[width=\figtwowidth]{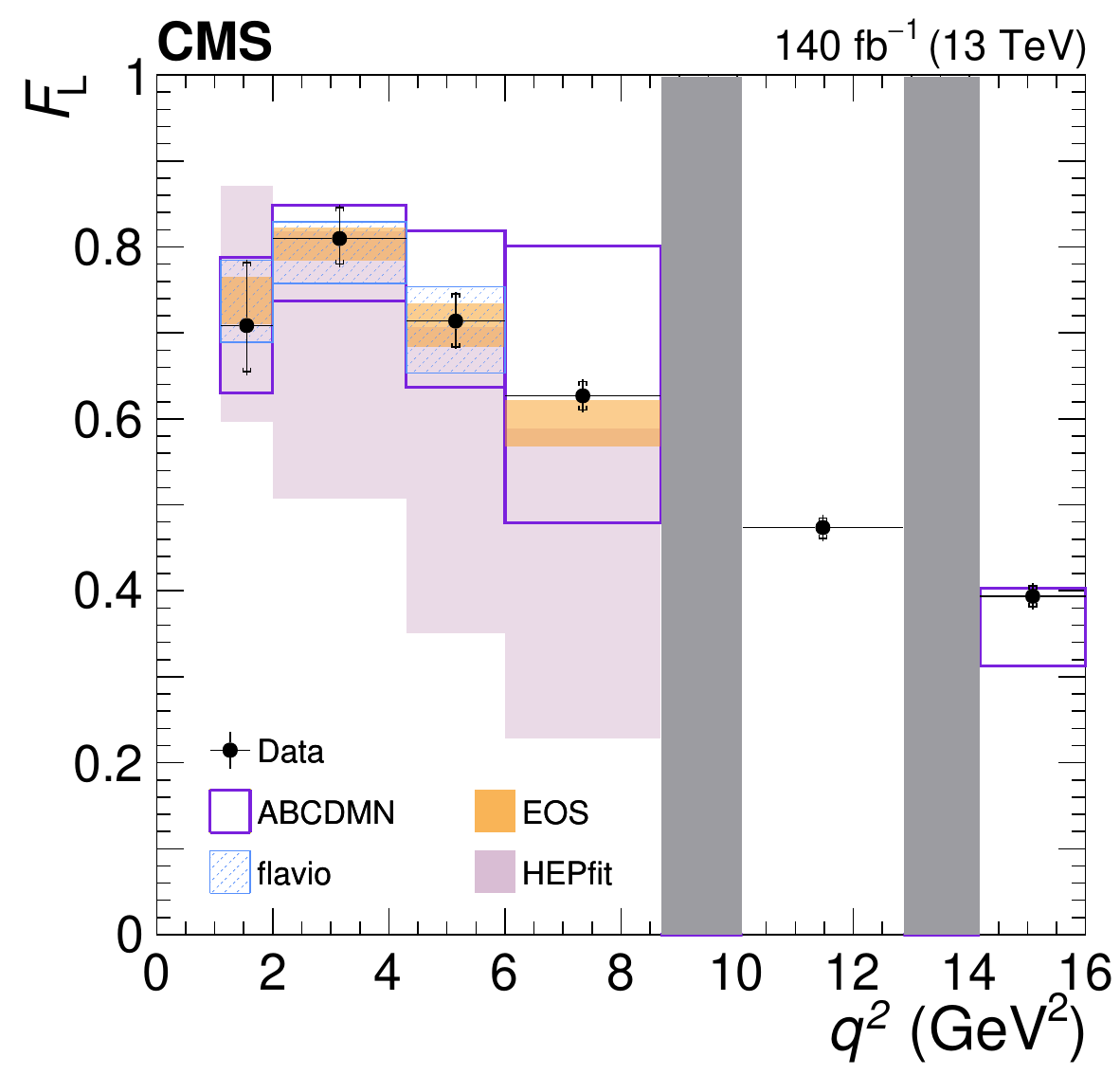}
    \includegraphics[width=\figtwowidth]{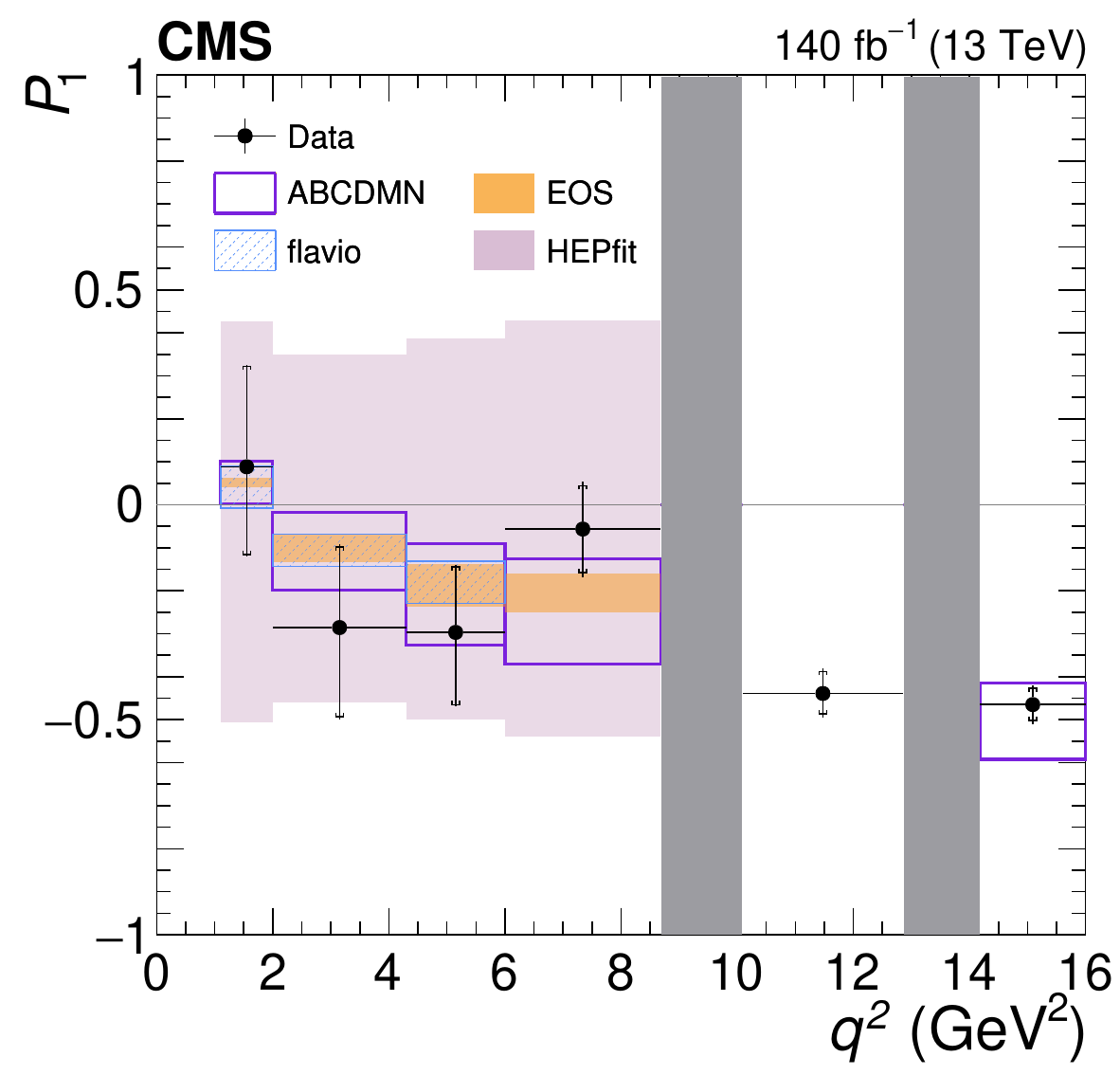}\\
    \includegraphics[width=\figtwowidth]{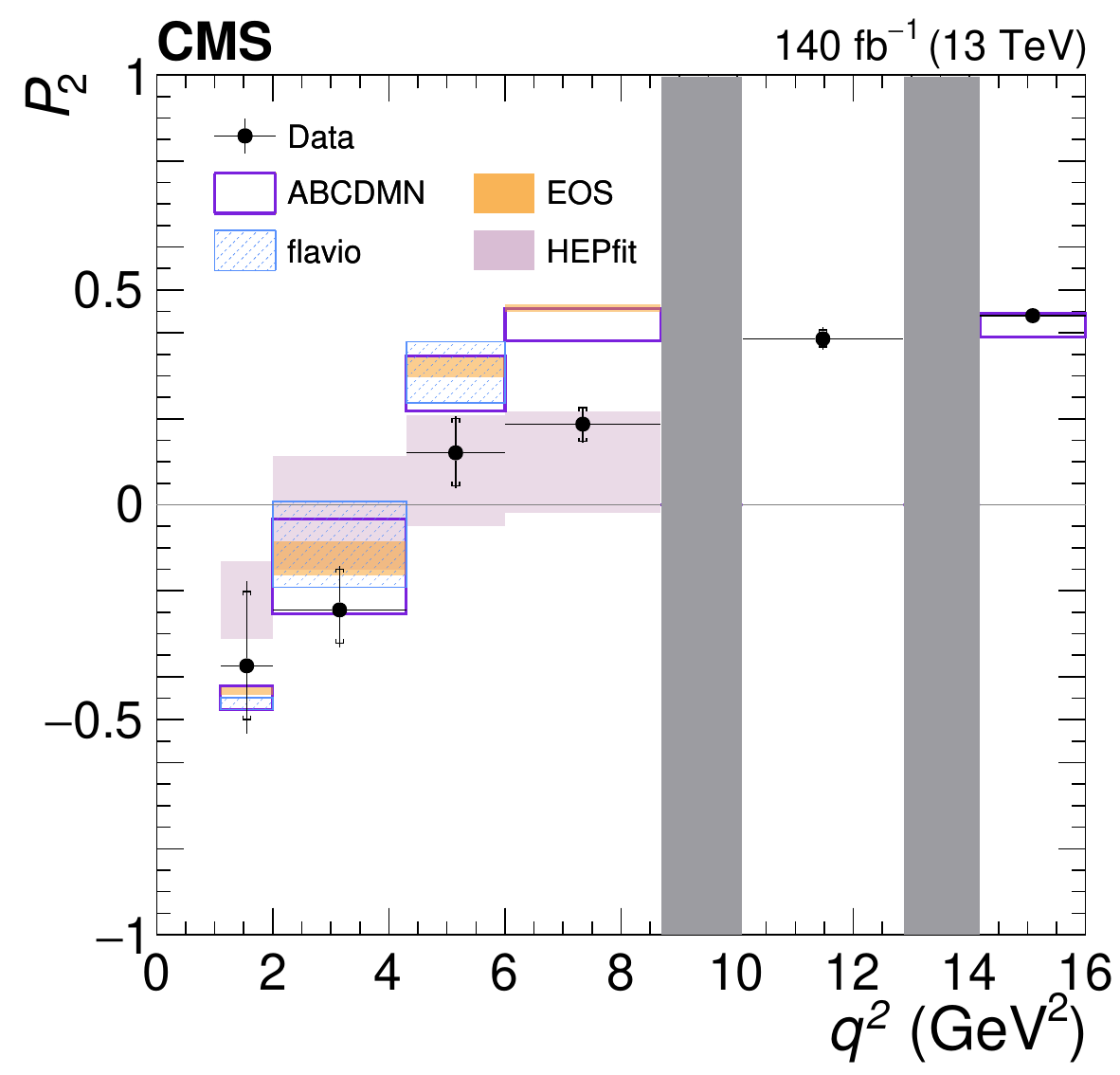}
    \includegraphics[width=\figtwowidth]{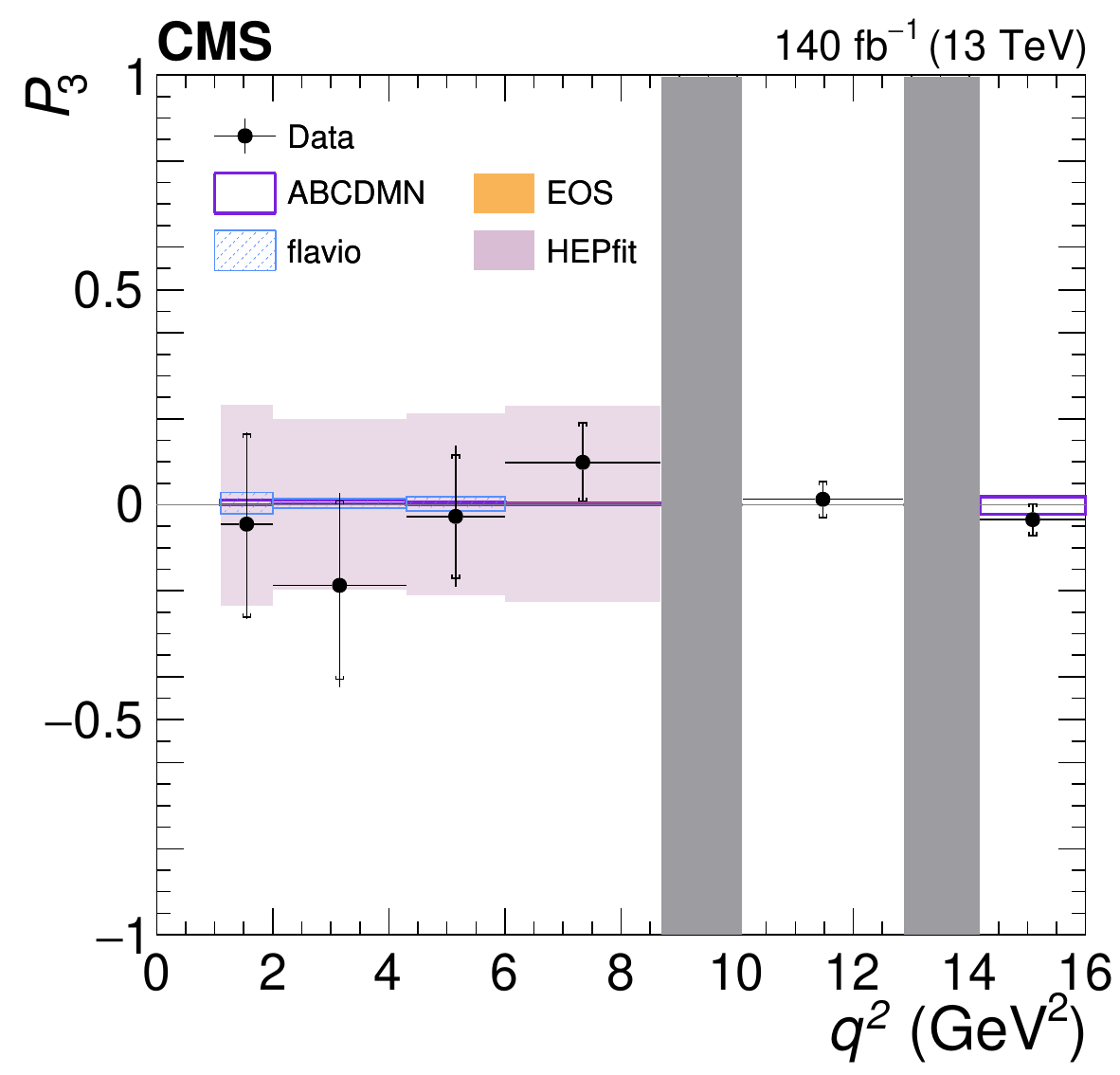}\\
    \includegraphics[width=\figtwowidth]{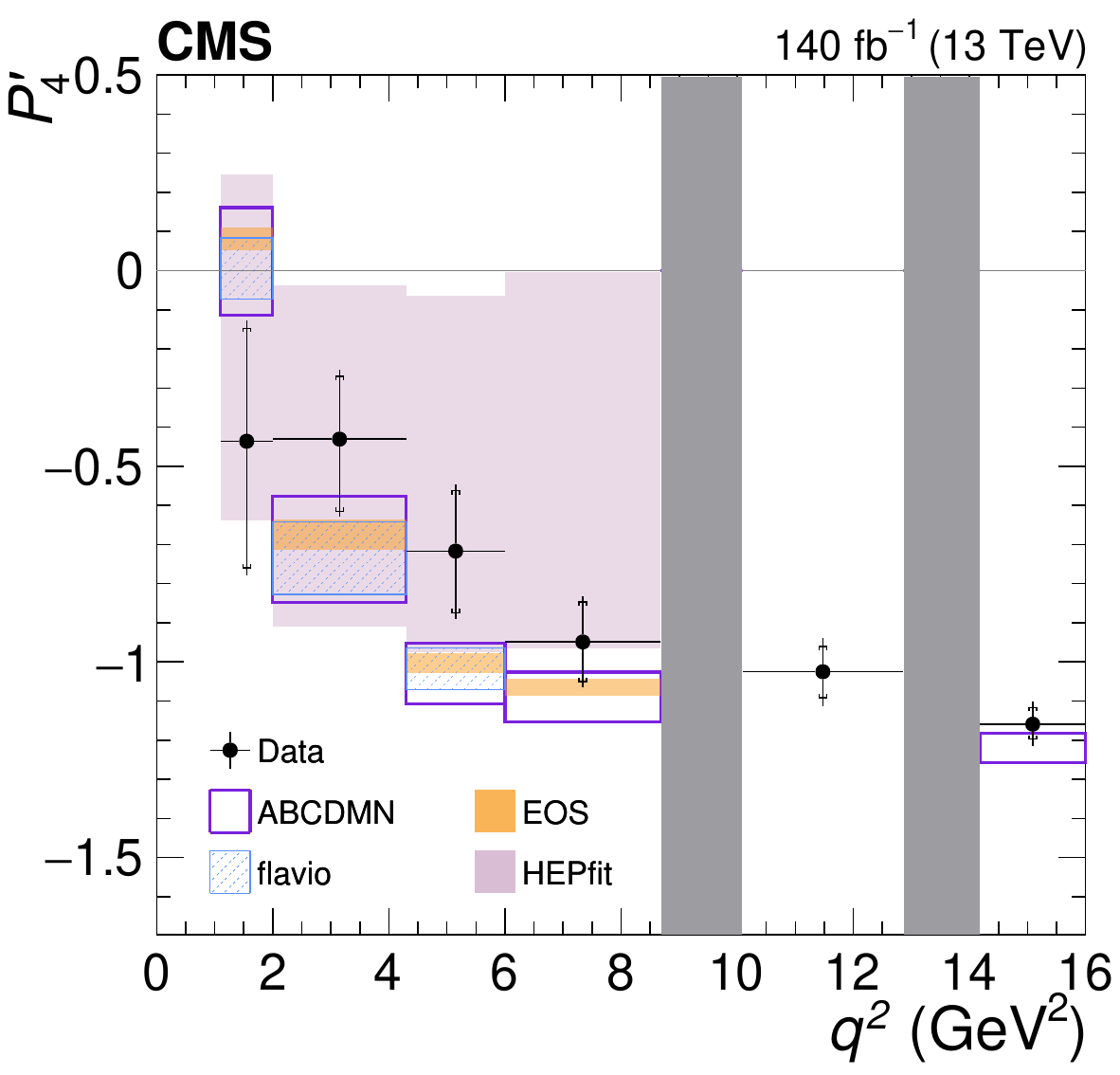}
    \includegraphics[width=\figtwowidth]{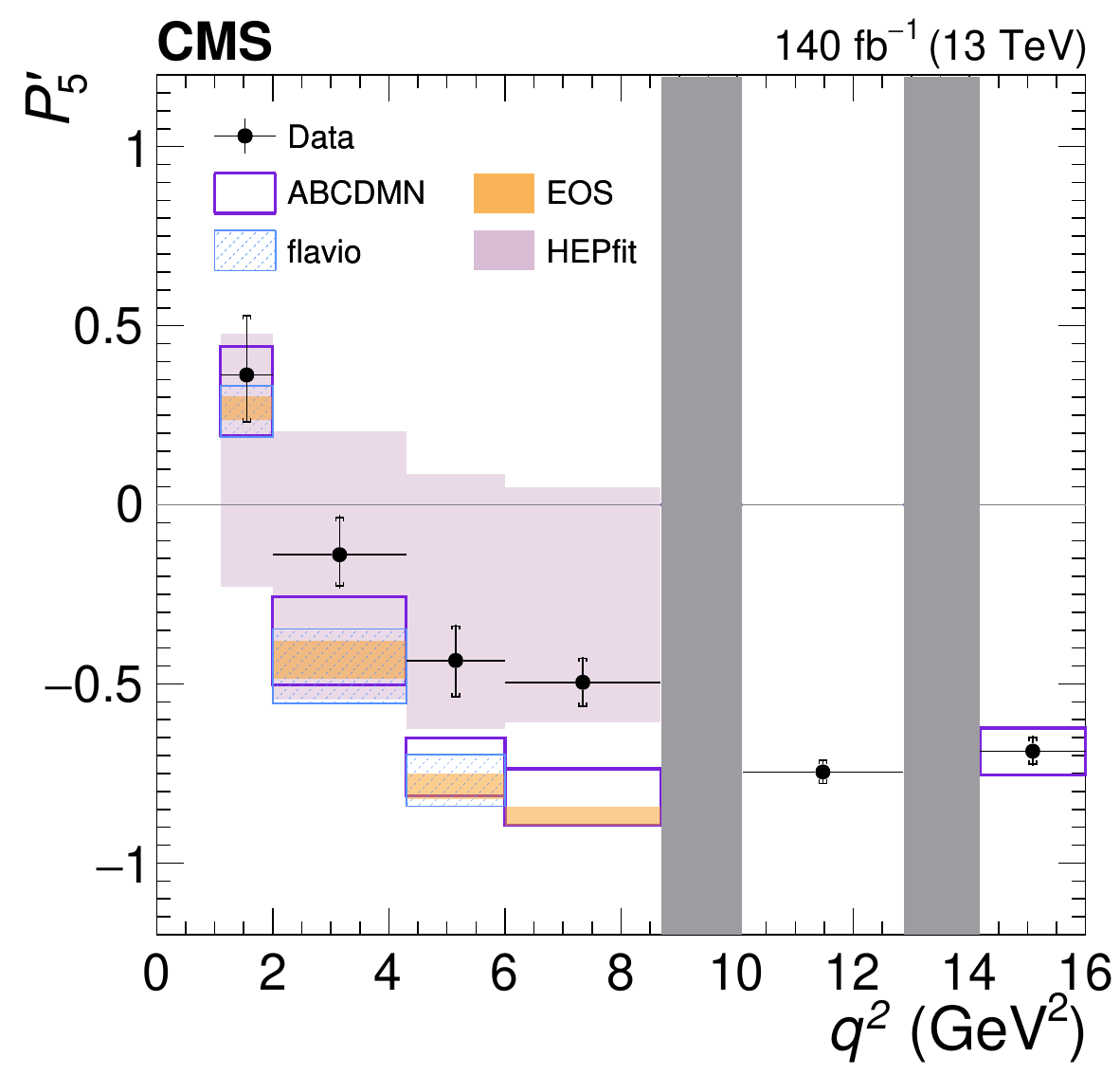}\\
    \includegraphics[width=\figtwowidth]{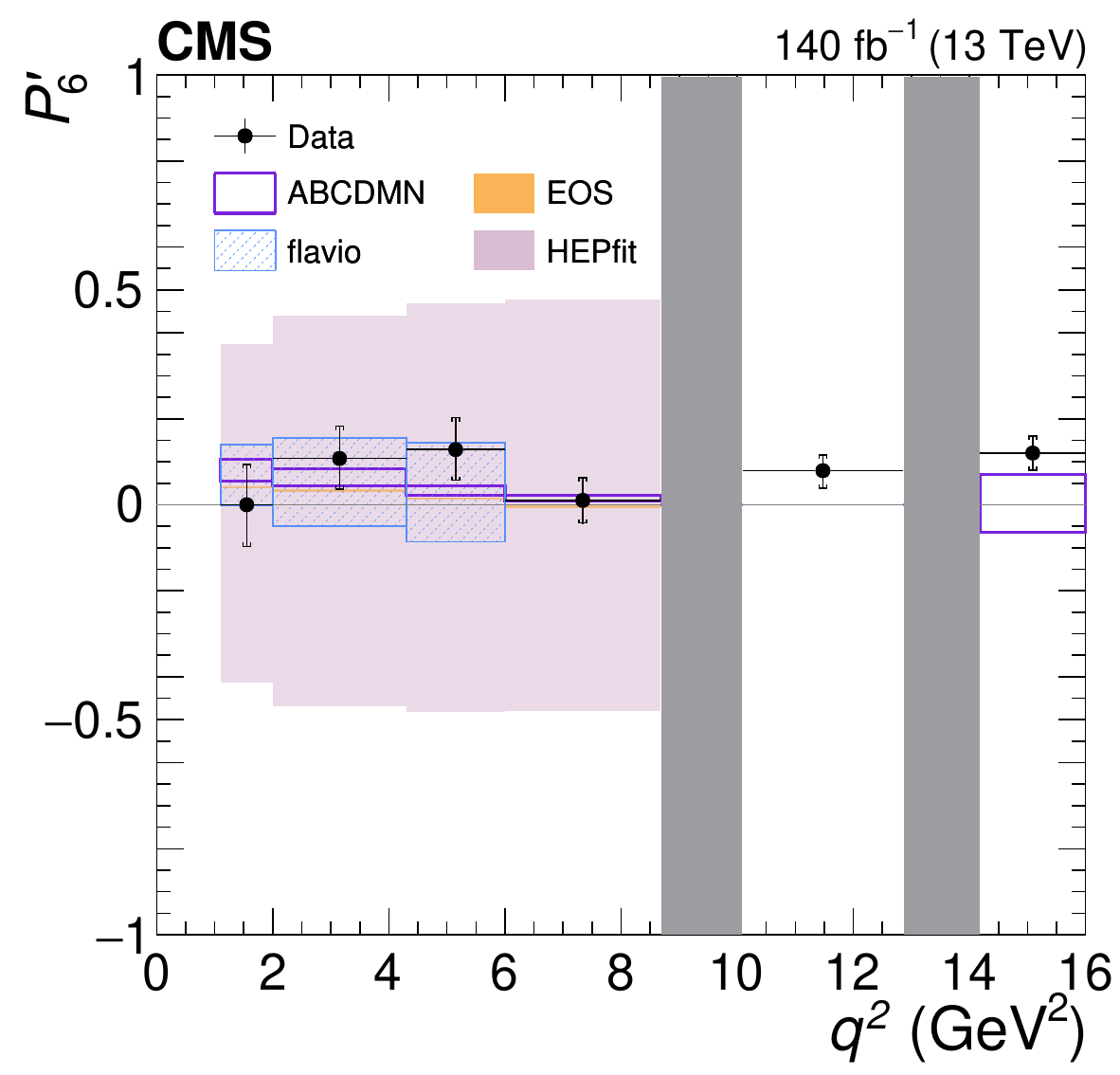}
    \includegraphics[width=\figtwowidth]{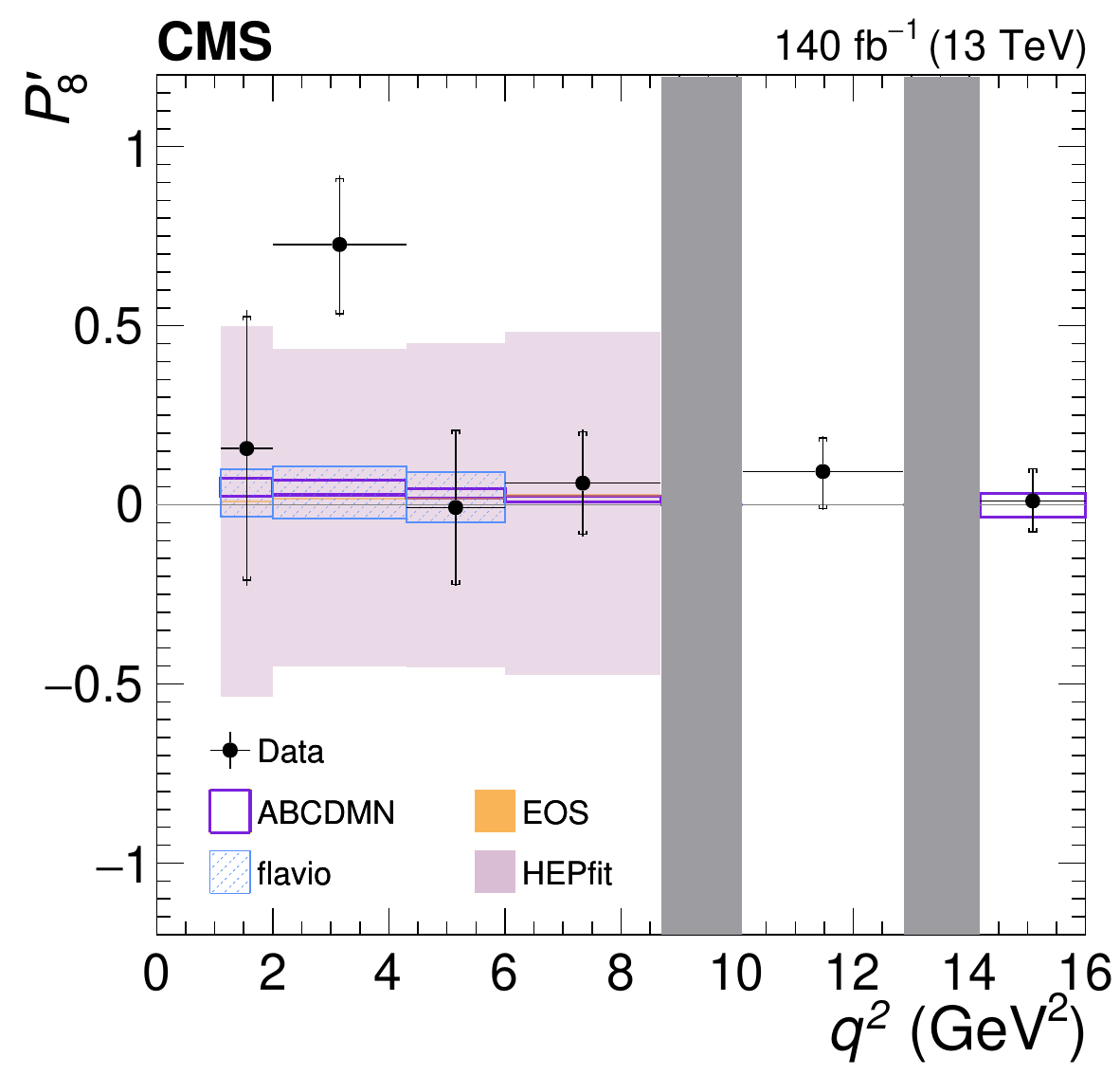}\\
    \caption{Measurements of the angular observables versus \qsq. The inner vertical bars represent the statistical uncertainties, while the outer vertical bars give the total uncertainties. The horizontal bars show the bin widths. The vertical shaded regions correspond to the \PJGy and \psit resonances.
    Predictions are shown, averaged in each bin, from Ref.~\cite{Alguero:2023jeh} (labeled ABCDMN), and the \texttt{EOS}~\cite{danny_van_dyk_2023_8340859},
    \texttt{flavio}~\cite{straub2018flavio}, and \texttt{HEPfit}~\cite{DeBlas:2019ehy} libraries.}
    \label{fig:unblind_step4_to_theory}
\end{figure*}

Table~\ref{tab:results} and Fig.~\ref{fig:unblind_step4_to_theory} provide the measured values of the \CP-averaged observables in bins of \qsq along with the corresponding uncertainties.
The size of the uncertainties varies between different $P_i^{(\prime)}$ observables, based on their dependence on \fl and, more generally, on the formula defining the observable.
Figure~\ref{fig:unblind_step4_to_theory} also contains four predictions based on the SM. The \qsq range of validity of the predictions varies depending on the methods used to estimate the form factors. 
The ABCDMN predictions are obtained using the theoretical framework discussed in Ref.~\cite{Alguero:2023jeh}, which combines local form factors derived from lattice quantum chromodynamics~\cite{Horgan:2013hoa,Horgan:2015vla} and light-cone sum rules~\cite{Gubernari:2018wyi}, and non-local form factors from Ref.~\cite{Gubernari:2020eft}.
The EOS predictions are obtained using the procedure described in Ref.~\cite{Gubernari:2022hxn} and the \texttt{EOS}~\cite{danny_van_dyk_2023_8340859} software package, using the same sources as ABCDMN for the form-factor calculations, but with a different approach to combining them. 
Additional predictions are calculated using the \texttt{flavio}~\cite{straub2018flavio} software package, which combines local form factors from lattice quantum chromodynamics~\cite{Horgan:2013hoa,Horgan:2015vla} and light-cone sum rules~\cite{Bharucha:2015bzk}, and quantum chromodynamics factorization for nonlocal form factors~\cite{Beneke:2004dp,Beneke:2001at}.
The predictions calculated with the \texttt{HEPfit}~\cite{DeBlas:2019ehy} software package adopt a conservative estimation of non-local hadronic matrix elements and their uncertainties, to account for a possible large impact from charm-loop penguin diagrams~\cite{Ciuchini:2015qxb,Ciuchini:2021smi}.

\begin{table*}[htp!]
  \topcaption{The measured \CP-averaged angular observables, in the corresponding \qsq bins. The first uncertainty is statistical and the second is systematic.
  \label{tab:results}}
  \centering
  \begin{tabular}{c c c c}
                   & $1.1 < q^2 < 2\GeV^2$                          & $\;\,2 < q^2 < 4.3\GeV^2$         & $\;4.3 < q^2 < 6\GeV^2$   \\\hline
   $\fl$\Tstrut & $\ph0.709\:\,^{ +\,0.073 }_{ -\,0.054 } \,\pm 0.021$            & $\ph0.810\,^{ +\,0.036 }_{ -\,0.030 } \pm 0.016$         & $\ph0.714\:\,^{ +\,0.032 }_{ -\,0.030 } \,\pm 0.012$   \\[.2cm]
   \pone        & $\ph0.09\phn\:\,^{ +\,0.23\phn }_{ -\,0.20\phn } \,\pm 0.04\phn$ & $-0.29\phn\,^{ +\,0.19\phn }_{ -\,0.21\phn } \pm 0.05\phn$ & $-0.30\phn\:\,^{ +\,0.15\phn }_{ -\,0.17\phn } \,\pm 0.04\phn$  \\[.2cm]
   \ptwo        & $-0.37\phn\:\,^{ +\,0.17\phn }_{ -\,0.13\phn } \,\pm 0.10\phn$   & $-0.244\,^{ +\,0.094 }_{ -\,0.077 } \pm 0.039$            & $\ph0.121\:\,^{ +\,0.080 }_{ -\,0.076 } \,\pm 0.030$   \\[.2cm]
   \pthree      & $-0.05\phn\:\,^{ +\,0.21\phn }_{ -\,0.22\phn } \,\pm 0.04\phn$   & $-0.19\phn\,^{ +\,0.20\phn }_{ -\,0.22\phn } \pm 0.09\phn$ & $-0.03\phn \pm 0.14 \pm 0.08\phn$  \\[.2cm]
   \pfourp      & $-0.44\phn\:\,^{ +\,0.29\phn }_{ -\,0.32\phn } \,\pm 0.11\phn$   & $-0.43\phn\,^{ +\,0.16\phn }_{ -\,0.19\phn } \pm 0.08\phn$ & $-0.72\phn\:\,^{ +\,0.15\phn }_{ -\,0.16\phn } \,\pm 0.07\phn$  \\[.2cm]
   \pfivep      & $\ph0.36\phn\:\,^{ +\,0.17\phn }_{ -\,0.13\phn } \,\pm 0.03\phn$ & $-0.14\phn\,^{ +\,0.10\phn }_{ -\,0.09\phn } \pm 0.04\phn$ & $-0.44\phn \pm 0.10 \pm 0.03\phn$  \\[.2cm]
   \psixp       & $\ph0.000\:\,^{ +\,0.094 }_{ -\,0.097 } \,\pm 0.021$            & $\ph0.108\,^{ +\,0.075 }_{ -\,0.071 } \pm 0.018$          & $\ph0.129\:\,^{ +\,0.074 }_{ -\,0.071 } \,\pm 0.011$   \\[.2cm]
   \peightp     & $\ph0.16\phn \pm 0.37 \pm 0.11\phn$                          & $\ph0.73\phn\,^{ +\,0.18\phn }_{ -\,0.19\phn } \pm 0.06\phn$ & $-0.01\phn \pm 0.22 \pm 0.04\phn$   \\
  \end{tabular}
  
 \bigskip
 \begin{tabular}{c c c c}
                & $6 < q^2 < 8.68\GeV^2$                 & $10.09 < q^2 < 12.86\GeV^2$            & $14.18 < q^2 < 16\GeV^2$        \\\hline
   $\fl$\Tstrut & $\ph0.627 \pm 0.016 \pm 0.011$                         & $\ph0.474\,^{ +\,0.011\! }_{ -\,0.013 } \pm 0.009$           & $\ph0.394 \pm 0.012 \pm 0.009$ \\[.2cm]
   \pone        & $-0.06\phn \pm 0.10\phn \pm 0.05\phn$                  & $-0.439\,^{ +\,0.051\! }_{ -\,0.047 } \pm 0.030$             & $-0.465 \pm 0.037 \pm 0.025$ \\[.2cm]
   \ptwo        & $\ph0.188\;\;^{ +\,0.039\! }_{ -\,0.040\! } \;\,\,\pm 0.014$ & $\ph0.386\,^{ +\,0.021\! }_{ -\,0.019 } \pm 0.018$           & $\ph0.440\;\;^{ +\,0.008\! }_{ -\,0.010\! } \;\,\,\pm 0.008$ \\[.2cm]
   \pthree      & $\ph0.099\;\;^{ +\,0.092\! }_{ -\,0.090\! } \;\,\,\pm 0.014$ & $\ph0.013\,^{ +\,0.041\! }_{ -\,0.043 } \pm 0.007$           & $-0.034\;\;^{ +\,0.037\! }_{ -\,0.038\! } \;\,\,\pm 0.010$ \\[.2cm]
   \pfourp      & $-0.95\phn \pm 0.10\phn \pm 0.06\phn$                  & $-1.025\,^{ +\,0.064\! }_{ -\,0.066 } \pm 0.059$             & $-1.159\;\;^{ +\,0.042\! }_{ -\,0.038\! } \;\,\,\pm 0.041$ \\[.2cm]
   \pfivep      & $-0.495 \pm 0.067 \pm 0.023$                           & $-0.746\,^{ +\,0.033\! }_{ -\,0.032 } \pm 0.014$             & $-0.688\;\;^{ +\,0.038\! }_{ -\,0.036\! } \;\,\,\pm 0.021$ \\[.2cm]
   \psixp       & $\ph0.010 \pm 0.052 \pm 0.016$                         & $\ph0.080\,^{ +\,0.037\! }_{ -\,0.041 } \pm 0.011$           & $\ph0.121\;\;^{ +\,0.040\! }_{ -\,0.039\! } \;\,\,\pm 0.011$ \\[.2cm]
   \peightp     & $\ph0.06\phn \pm 0.14\phn \pm 0.04\phn$                & $\ph0.09\phn\,^{ +\,0.09\phn\! }_{ -\,0.10\phn } \pm 0.03\phn$ & $\ph0.011\;\;^{ +\,0.089\! }_{ -\,0.086\! } \;\,\,\pm 0.022$ \\
  \end{tabular}
\end{table*}

\begin{figure}[hbtp!]
  \centering
    \includegraphics[width=0.45\textwidth]{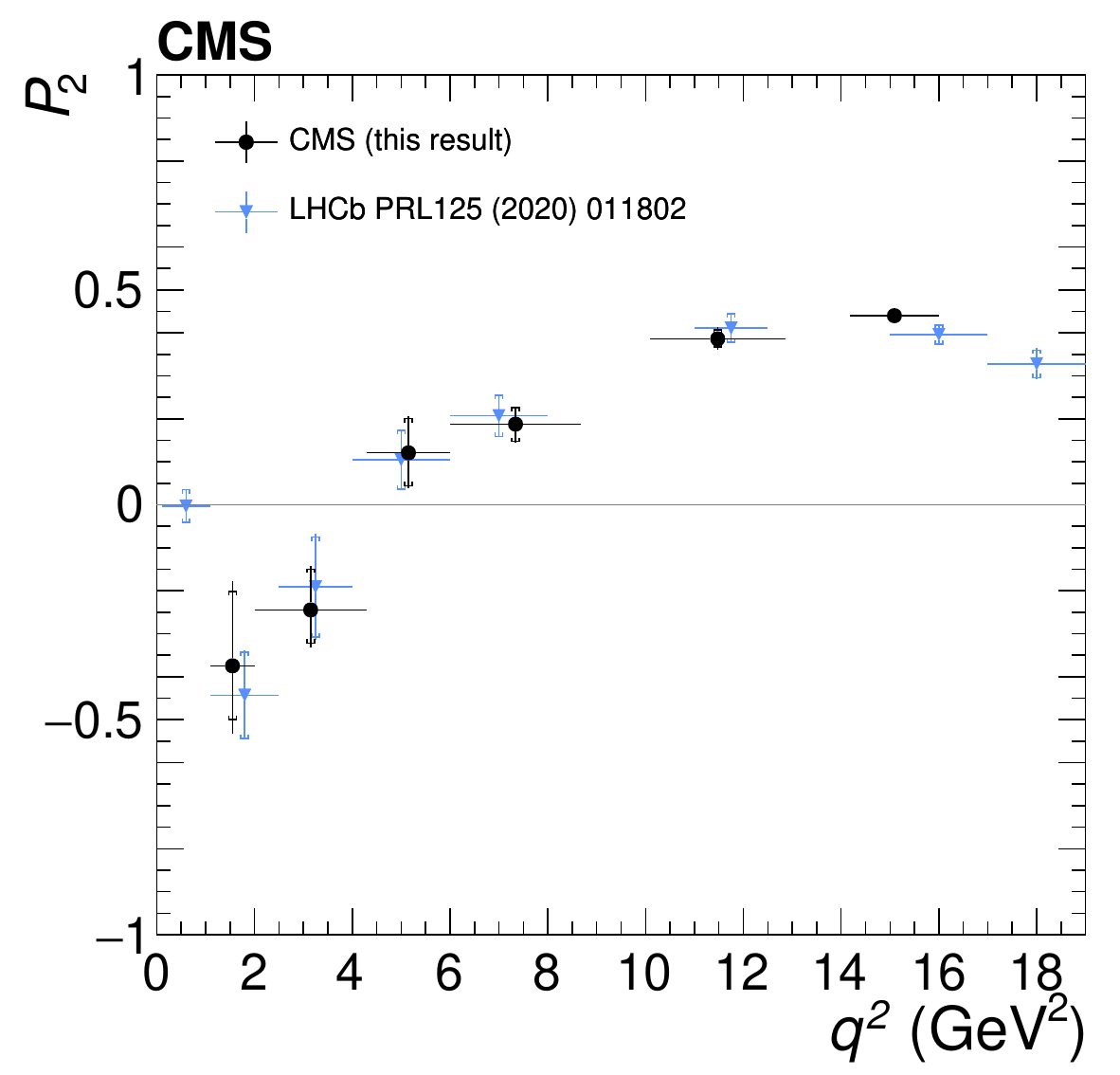}
    \includegraphics[width=0.45\textwidth]{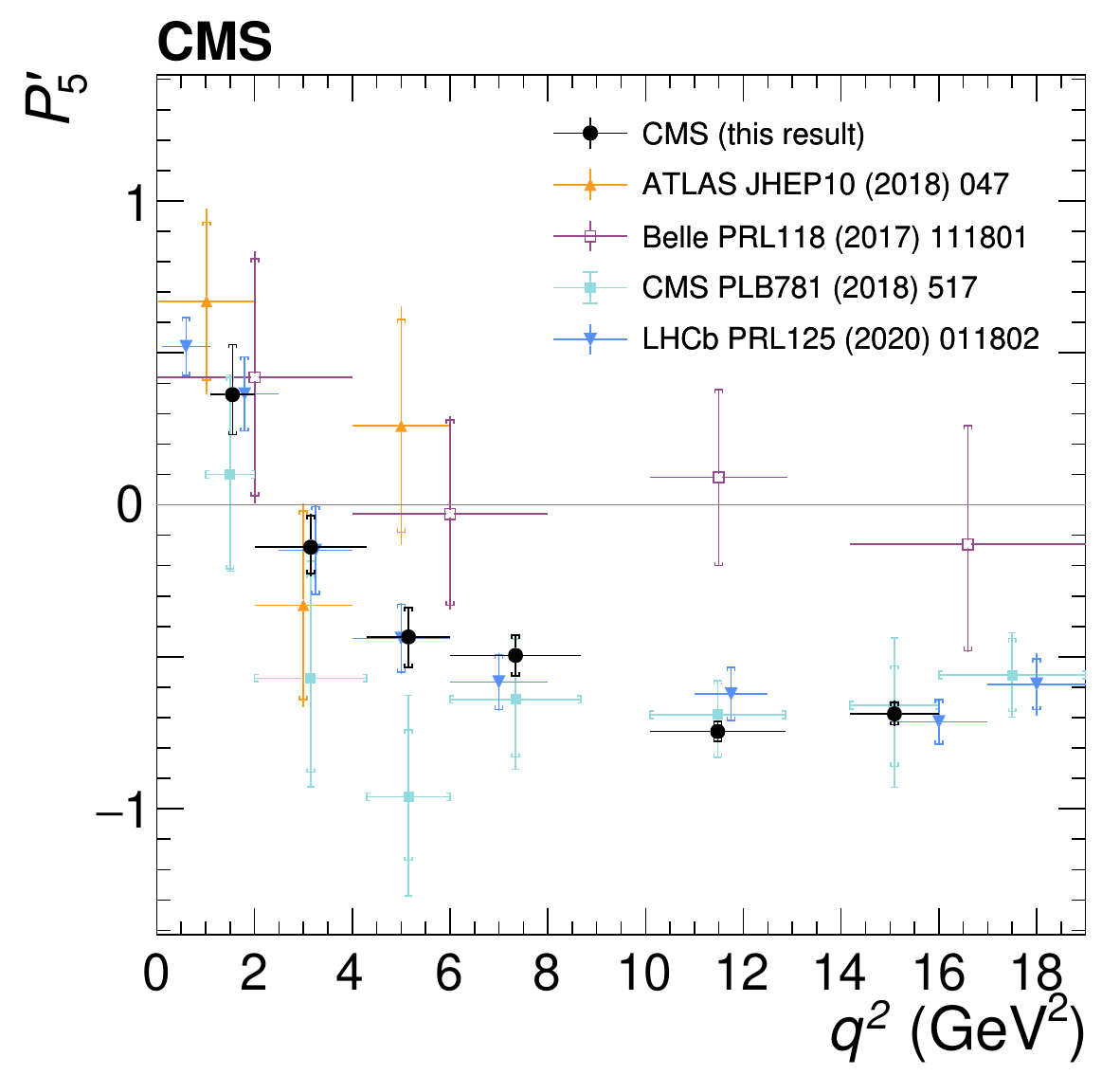}
    \caption{Comparison of current measurements of \ptwo and \pfivep to a previous \ptwo measurement by LHCb~\cite{Aaij:2020nrf} and previous \pfivep results from ATLAS~\cite{ATLAS:2018gqc}, Belle~\cite{Wehle:2016yoi}, CMS~\cite{Sirunyan:2017dhj}, and LHCb~\cite{Aaij:2020nrf}.
    The inner vertical bars represent the statistical uncertainties, while the outer vertical bars give the total uncertainties. The horizontal bars show the bin widths. 
}
    \label{fig:unblind_step4_to_exp}
\end{figure}

The HEPfit predictions, given the large uncertainties, are compatible with the measurements.
The EOS and ABCDMN predictions, while showing compatibility with the analysis results for most of the observables, present tensions with the measured values of \ptwo and \pfivep, in the region of \qsq below the \PJGy resonance. 
The deviation for \pfivep is at a level of 3.2 (2.2) and 4.9 (3.0) standard deviations for EOS (ABCDMN) predictions in the regions $4.3 < \qsq < 6\GeV^2$ and $6 < \qsq < 8.68\GeV^2$, respectively, while for \ptwo the difference corresponds to 2.2 (1.5) and 6.4 (4.1) standard deviations for EOS (ABCDMN) predictions, in the same \qsq ranges. 
The flavio predictions show similar tensions for the \pfivep parameter, at the level of 2.6 standard deviations in the regions $4.3 < \qsq < 6\GeV^2$.
The value of \peightp in the second bin of \qsq disagrees with the predictions, with a significance that ranges between 1.5 and 3.5 standard deviations.
The current measurements are compatible with those presented in previous analyses from the CMS Collaboration~\cite{CMS:2015bcy,Sirunyan:2017dhj} and other experiments~\cite{Aaij:2015oid,Wehle:2016yoi,Aaij:2020nrf,ATLAS:2018gqc}, as shown for \ptwo and \pfivep in Fig.~\ref{fig:unblind_step4_to_exp}.
The statistical uncertainty has significantly improved compared to previous CMS publications.

\section{Summary}
In summary, the study of the full angular distribution of the flavor-changing neutral-current \bzKstarmm decay has been performed using 140\fbinv of proton-proton collision data recorded by the CMS detector at the LHC at $\sqrt{s}= 13\TeV$.
A complete set of observables has been measured via unbinned maximum likelihood fits to the \PBz candidate mass and angular variables, in bins of the squared invariant mass of the dimuon system ranging from 1.1 to $16\GeV^2$.
The measurements are compared to a variety of predictions based on the standard model, with tension in a few of the angular observables seen for some of the predictions, as is also reported by other experiments.
These results are among the most precise experimental measurements of the angular observables of this decay, and provide a valuable contribution to the understanding of the \bsll processes.

\begin{acknowledgments}
We congratulate our colleagues in the CERN accelerator departments for the excellent performance of the LHC and thank the technical and administrative staffs at CERN and at other CMS institutes for their contributions to the success of the CMS effort. In addition, we gratefully acknowledge the computing centres and personnel of the Worldwide LHC Computing Grid and other centres for delivering so effectively the computing infrastructure essential to our analyses. Finally, we acknowledge the enduring support for the construction and operation of the LHC, the CMS detector, and the supporting computing infrastructure provided by the following funding agencies: SC (Armenia), BMBWF and FWF (Austria); FNRS and FWO (Belgium); CNPq, CAPES, FAPERJ, FAPERGS, and FAPESP (Brazil); MES and BNSF (Bulgaria); CERN; CAS, MoST, and NSFC (China); MINCIENCIAS (Colombia); MSES and CSF (Croatia); RIF (Cyprus); SENESCYT (Ecuador); ERC PRG, RVTT3 and MoER TK202 (Estonia); Academy of Finland, MEC, and HIP (Finland); CEA and CNRS/IN2P3 (France); SRNSF (Georgia); BMBF, DFG, and HGF (Germany); GSRI (Greece); NKFIH (Hungary); DAE and DST (India); IPM (Iran); SFI (Ireland); INFN (Italy); MSIP and NRF (Republic of Korea); MES (Latvia); LMTLT (Lithuania); MOE and UM (Malaysia); BUAP, CINVESTAV, CONACYT, LNS, SEP, and UASLP-FAI (Mexico); MOS (Montenegro); MBIE (New Zealand); PAEC (Pakistan); MES and NSC (Poland); FCT (Portugal); MESTD (Serbia); MCIN/AEI and PCTI (Spain); MOSTR (Sri Lanka); Swiss Funding Agencies (Switzerland); MST (Taipei); MHESI and NSTDA (Thailand); TUBITAK and TENMAK (Turkey); NASU (Ukraine); STFC (United Kingdom); DOE and NSF (USA).

\hyphenation{Rachada-pisek} Individuals have received support from the Marie-Curie programme and the European Research Council and Horizon 2020 Grant, contract Nos.\ 675440, 724704, 752730, 758316, 765710, 824093, 101115353, 101002207, and COST Action CA16108 (European Union); the Leventis Foundation; the Alfred P.\ Sloan Foundation; the Alexander von Humboldt Foundation; the Science Committee, project no. 22rl-037 (Armenia); the Belgian Federal Science Policy Office; the Fonds pour la Formation \`a la Recherche dans l'Industrie et dans l'Agriculture (FRIA-Belgium); the F.R.S.-FNRS and FWO (Belgium) under the ``Excellence of Science -- EOS" -- be.h project n.\ 30820817; the Beijing Municipal Science \& Technology Commission, No. Z191100007219010 and Fundamental Research Funds for the Central Universities (China); the Ministry of Education, Youth and Sports (MEYS) of the Czech Republic; the Shota Rustaveli National Science Foundation, grant FR-22-985 (Georgia); the Deutsche Forschungsgemeinschaft (DFG), among others, under Germany's Excellence Strategy -- EXC 2121 ``Quantum Universe" -- 390833306, and under project number 400140256 - GRK2497; the Hellenic Foundation for Research and Innovation (HFRI), Project Number 2288 (Greece); the Hungarian Academy of Sciences, the New National Excellence Program - \'UNKP, the NKFIH research grants K 131991, K 133046, K 138136, K 143460, K 143477, K 146913, K 146914, K 147048, 2020-2.2.1-ED-2021-00181, TKP2021-NKTA-64, and 2021-4.1.2-NEMZ\_KI (Hungary); the Council of Science and Industrial Research, India; ICSC -- National Research Centre for High Performance Computing, Big Data and Quantum Computing and FAIR -- Future Artificial Intelligence Research, funded by the NextGenerationEU program (Italy); the Latvian Council of Science; the Ministry of Education and Science, project no. 2022/WK/14, and the National Science Center, contracts Opus 2021/41/B/ST2/01369 and 2021/43/B/ST2/01552 (Poland); the Funda\c{c}\~ao para a Ci\^encia e a Tecnologia, grant CEECIND/01334/2018 (Portugal); the National Priorities Research Program by Qatar National Research Fund; MCIN/AEI/10.13039/501100011033, ERDF ``a way of making Europe", and the Programa Estatal de Fomento de la Investigaci{\'o}n Cient{\'i}fica y T{\'e}cnica de Excelencia Mar\'{\i}a de Maeztu, grant MDM-2017-0765 and Programa Severo Ochoa del Principado de Asturias (Spain); the Chulalongkorn Academic into Its 2nd Century Project Advancement Project, and the National Science, Research and Innovation Fund via the Program Management Unit for Human Resources \& Institutional Development, Research and Innovation, grant B39G670016 (Thailand); the Kavli Foundation; the Nvidia Corporation; the SuperMicro Corporation; the Welch Foundation, contract C-1845; and the Weston Havens Foundation (USA).
\end{acknowledgments}

\bibliography{auto_generated} 
\cleardoublepage \appendix\section{The CMS Collaboration \label{app:collab}}\begin{sloppypar}\hyphenpenalty=5000\widowpenalty=500\clubpenalty=5000\input{BPH-21-002-public-authorlist.tex}\end{sloppypar}
\end{document}

%% file: BPH-21-002-public-authorlist.tex
\cmsinstitute{Yerevan Physics Institute, Yerevan, Armenia}
{\tolerance=6000
A.~Hayrapetyan, A.~Tumasyan\cmsAuthorMark{1}\cmsorcid{0009-0000-0684-6742}
\par}
\cmsinstitute{Institut f\"{u}r Hochenergiephysik, Vienna, Austria}
{\tolerance=6000
W.~Adam\cmsorcid{0000-0001-9099-4341}, J.W.~Andrejkovic, L.~Benato\cmsorcid{0000-0001-5135-7489}, T.~Bergauer\cmsorcid{0000-0002-5786-0293}, S.~Chatterjee\cmsorcid{0000-0003-2660-0349}, K.~Damanakis\cmsorcid{0000-0001-5389-2872}, M.~Dragicevic\cmsorcid{0000-0003-1967-6783}, P.S.~Hussain\cmsorcid{0000-0002-4825-5278}, M.~Jeitler\cmsAuthorMark{2}\cmsorcid{0000-0002-5141-9560}, N.~Krammer\cmsorcid{0000-0002-0548-0985}, A.~Li\cmsorcid{0000-0002-4547-116X}, D.~Liko\cmsorcid{0000-0002-3380-473X}, I.~Mikulec\cmsorcid{0000-0003-0385-2746}, J.~Schieck\cmsAuthorMark{2}\cmsorcid{0000-0002-1058-8093}, R.~Sch\"{o}fbeck\cmsAuthorMark{2}\cmsorcid{0000-0002-2332-8784}, D.~Schwarz\cmsorcid{0000-0002-3821-7331}, M.~Sonawane\cmsorcid{0000-0003-0510-7010}, W.~Waltenberger\cmsorcid{0000-0002-6215-7228}, C.-E.~Wulz\cmsAuthorMark{2}\cmsorcid{0000-0001-9226-5812}
% ...
\par}
\vskip\cmsinstskip
\dag:~Deceased\\
$^{1}$Also at Yerevan State University, Yerevan, Armenia\\
$^{2}$Also at TU Wien, Vienna, Austria\\
% ...
